\documentclass[aps,prd,showpacs,floats,letterpaper,floatfix,
groupedaddress,superscriptaddress
,nofootinbib
,eqsecnum
,twocolumn
]{revtex4}
\usepackage{amsmath}
\usepackage{amsfonts}
\usepackage{bm}
\usepackage{graphicx}
  \newcommand{\mnras}{Mon. Not. R. Astron. Soc.}

   \newcommand{\aap}{Astron. Astrophys.}
   
   \newcommand{\aj}{Astron. J.}

\def\v1v2{{\bf v}_1 \cdot {\bf v}_2}
\def\bx{\bm{x}}
\allowdisplaybreaks

\begin{document}

\title{Higher-order effects in the dynamics of hierarchical triple systems.  Quadrupole-squared terms}

\author{
Clifford M.~Will} \email{cmw@phys.ufl.edu}
\affiliation{Department of Physics, University of Florida, Gainesville, Florida 32611, USA}
\affiliation{GReCO, Institut d'Astrophysique de Paris, CNRS,\\ 
Universit\'e Pierre et Marie Curie, 98 bis Boulevard Arago, 75014 Paris, France}

\date{\today}

\begin{abstract}
We analyze the secular evolution of hierarchical triple systems to second-order in the quadrupolar perturbation induced on the inner binary by the distant third body.   The Newtonian three-body equations of motion, expanded in powers of the ratio of semimajor axes $a/A$,  become a pair of effective one-body Keplerian equations of motion, perturbed by a sequence of multipolar perturbations, denoted quadrupole, $O[(a/A)^3]$, octupole, $O[(a/A)^4]$, and so on.  In the Lagrange planetary equations for the evolution of the instantaneous orbital elements, second-order effects arise from obtaining the first-order solution for each element, consisting of a constant (or slowly varying) piece and an oscillatory perturbative piece, and reinserting it back into the equations to obtain a second-order solution.  After an average over the two orbital timescales to obtain long-term evolutions, these second-order quadrupole ($Q^2$) terms would be expected to produce effects of order $(a/A)^6$. However we find that the orbital average actually enhances the second-order terms by a factor of the ratio of the outer to the inner orbital periods, $ \sim (A/a)^{3/2}$.  For systems with a low-mass third body, the $Q^2$ effects are small, but for systems with a comparable-mass or very massive third body, such as a Sun-Jupiter system orbiting a solar-mass star, or a $100 \, M_\odot$ binary system orbiting a $10^6 \, M_\odot$ massive black hole, the $Q^2$ effects can completely suppress flips of the inner orbit from prograde to retrograde and back that occur in the first-order solutions.   These results are in complete agreement with those of Luo, Katz and Dong, derived using a ``Corrected Double-Averaging'' method.

\end{abstract}

\maketitle

\section{Introduction and summary}
\label{sec:intro}

The hierarchical three-body problem is rich in interesting dynamics as well as astrophysical applications.  It is a special case of the general three-body problem, in which an inner binary system is in orbit with a third body at a distance large compared to the average separation within the inner binary.   With suitable conditions on the masses and separations, the problem can be formulated using perturbation theory.  At the lowest order, the orbits of the inner binary and of the third body relative to the center of mass of the inner binary are standard exact solutions of the Newtonian two-body problem.

One then determines the perturbations to this system by expanding Newton's equations in powers of the parameter $\epsilon = a/A \ll 1$, where $a$ is the semimajor axis of the inner  binary, and $A$ is the semimajor axis of the outer ``binary''.  The results include perturbations of the inner binary due to the third body and perturbations of the third body due to the finite extent of the inner binary's mass distribution.  One then obtains a sequence of perturbing terms in the equations of motion involving progressively higher powers of $\epsilon$.  Relative to the dominant Newtonian two-body acceleration, for the inner binary, these terms have the amplitudes $\alpha \epsilon^3$, $\alpha \epsilon^4$, $\alpha \epsilon^5$, $\alpha \epsilon^6$, and so on, while for the outer binary, they have the amplitudes $\eta \epsilon^2$, $\eta \epsilon^3$, $\eta \epsilon^4$, $\eta \epsilon^5$, and so on, where $\alpha = m_3/(m_1+m_2)$ and $\eta = m_1m_2/(m_1+m_2)^2$.   Each level of the expansion is assigned a specific name: quadrupole, octupole, hexadecapole, dotriocontopole, etc.  The masses are arbitrary, apart from the constraint that $\alpha \epsilon^3$ be sufficiently small that the dominant quadrupole term for the inner binary be a suitably small perturbation.  After averaging over the short orbital timescales, one obtains equations for the long-term evolution of the orbital elements such as eccentricity and inclination for each orbit.

Hierarchical triples have been enshrined in physics and astronomy history. Notable examples include the Earth-Moon system perturbed by the Sun, studied by Newton, Clairaut and many others, and the Sun-Mercury system perturbed by each of the other planets, studied by Le Verrier and made whole by Einstein.  In the 1960s, working at the leading quadrupole order of approximation, Lidov and Kozai  \cite{1962P&SS....9..719L,1962AJ.....67..591K} found the remarkable oscillations involving an interchange between the eccentricity of the two-body inner orbit and its inclination relative to the plane of the third body.  The Kozai-Lidov oscillations were derived assuming a circular outer orbit, but generalizing to eccentric outer orbits and adding octupole terms,  Naoz et al. \cite{2011Natur.473..187N,2013MNRAS.431.2155N}, following up earlier theoretical work \cite{1999MNRAS.304..720K,2000ApJ...535..385F,2002ApJ...578..775B} found the possibility of complete ``flips'' of the inner orbital plane, accompanied by excursions to extreme values of its eccentricity, providing a possible explanation of retrograde ``hot Jupiters'' in some exoplanet systems.

Authors have explored even higher multipole terms in the perturbation expansion, partly in search of interesting new phenomena, and partly to obtain equations that would enable more accurate long-term evolutions of hierarchical triple systems \cite{2010A&A...522A..60L,2015MNRAS.452.3610A,2016MNRAS.459.2827H,2016CeMDA.124...73C,2017PhRvD..96b3017W}.  

In our work obtaining the equations to hexadecapole order \cite{2017PhRvD..96b3017W} ($O(\alpha \epsilon^5)$ for the inner orbit and $O(\eta \epsilon^4)$ for the outer orbit), we used the approach of ``osculating orbit elements'' whereby each of the orbits is characterized by its instantaneous semimajor axis  and eccentricity, its inclination and angle of ascending node relative to a reference coordinate system, and its angle of pericenter measured from the ascending node.  The equations of motion for the two orbits can then be rewritten as the ``Lagrange planetary'' equations for the orbit elements, which take the generic form
\begin{equation}
\frac{dX^\alpha}{dt} = Q^\alpha (X^\beta, t) \,,
\label{eq:1}
\end{equation}
where  
$X^\alpha$ denotes orbit elements of the inner and outer binary.   We then carried out the conventional average over an orbit of both the inner binary and the outer binary holding the orbit elements fixed, arriving at equations for the secular changes in the orbit elements.   

In contemplating extending our work to dotriocontopole order, namely $O(\alpha \epsilon^6)$ for the inner orbit, we realized that there would be contributions to the evolution equations for the orbit elements at the same order as dotriocontopole, but that would not be revealed by the simple averaging process described above.  Instead, one must take into account that each osculating orbit element actually consists of a constant (or slowly varying) part and a part that has variations on the orbital timescales, induced by the quadrupole perturbations.  That oscillatory piece would have an amplitude $\alpha \epsilon^3$.  Substituting that first-order solution back into the Lagrange planetary equations and averaging again would in general lead to a second-order contribution with amplitude $\alpha^2 \epsilon^6$, the same as dotriocontopole order, apart from an additional factor of $\alpha$.  We call these ``quadrupole-quadrupole'', or $Q^2$ contributions.  This simply reflects the fact that, while the equations of motion are linear in the multipoles, the solutions of the equations are not, simply because the multipolar perturbations depend on the orbital variables, which themselves are perturbed. 

However, when dealing with second-order perturbations in the Lagrange planetary equations, we must revisit the procedure for the double average over the two orbital timescales.  Because the hierarchical assumption requires $\epsilon \ll 1$ and the perturbative assumption requires $\alpha \epsilon^3 \ll 1$, the ratio of the inner to the outer  orbital period is automatically small, i.e.
\begin{align}
\frac{P_{\rm in}}{P_{\rm out}} = (1+\alpha)^{1/2} \epsilon^{3/2} \ll 1
\end{align} 
In first-order perturbation theory, where the orbit elements are treated as constants, all the terms on the right-hand side of the planetary equations are the product of periodic functions that vary on the short (inner) orbital timescale with periodic functions that vary on the long (outer) orbital timescale.  It can be shown that the average of such products is the product of the separate averages, up to corrections of order $(P_{\rm in}/P_{\rm out})^2$.  Physically this is equivalent to holding the slowly moving outer body fixed while averaging over an inner orbit, then averaging over the outer body's orbit.  This averaging procedure is often called the ``secular approximation''. 

But at second order, we no longer have simple products of periodic functions, because the first-order perturbations of each orbit element that have been reinserted into the planetary equations are {\em integrals} of products of periodic functions, because they are, after all, solutions of the first-order equations (\ref{eq:1}).
Averaging products of periodic functions multiplied by  these integrals yields two types of terms.  One type is the expected second-order term, of order $(\alpha \epsilon^3)^2$, as we discussed above.  These would be comparable to dotriocontopole terms apart from the extra $\alpha$ factor.    However the averaging yields a second type of term that is larger than this by the ratio  $P_{\rm out}/P_{\rm in}$.  This term leads to a contribution to the evolution equations for the inner orbit elements of order
$\alpha^2 (1+\alpha)^{-1/2} \epsilon^{9/2}$.    These contributions are ``midway'' between octupole ($\alpha \epsilon^4$) and hexadecapole ($\alpha \epsilon^5$) terms, and for high outer-mass systems ($\alpha \gg 1$), they could actually dominate octupole terms.  In this paper, 
we will focus entirely on these dominant $Q^2$ contributions, and ignore the terms that are of dotriocontopole order.

We solve the Lagrange planetary equations for the osculating orbit elements using a two-timescale analysis, in which the short timescale is defined by the two orbital periods, and the long timescale is associated with the perturbations \cite{1978amms.book.....B,1990PhRvD..42.1123L,2004PhRvD..69j4021M,2008PhRvD..78f4028H,2017PhRvD..95f4003W}.  This method is well suited to implementing higher-order perturbation theory on systems like the Lagrange planetary equations.   It has been used effectively to derive orbit evolution equations for the two-body problem to high orders in the post-Newtonian approximation of general relativity \cite{2004PhRvD..69j4021M,2017PhRvD..95f4003W,tuckerwill}, and to analyse ``post-Newtonian cross-terms'' in hierarchical triples, generated by post-Newtonian corrections to the perturbing terms in the equations of motion  \cite{2020PhRvD.102f4033L,willcrossterms}.  Combining the resulting $Q^2$ terms with the first-order contributions through hexadecapole order, we evolve the equations numerically for interesting astrophysical cases.  We also include the leading general relativistic pericenter precessions for each orbit.

\begin{figure}[t]
\begin{center}

\includegraphics[width=3.4in]{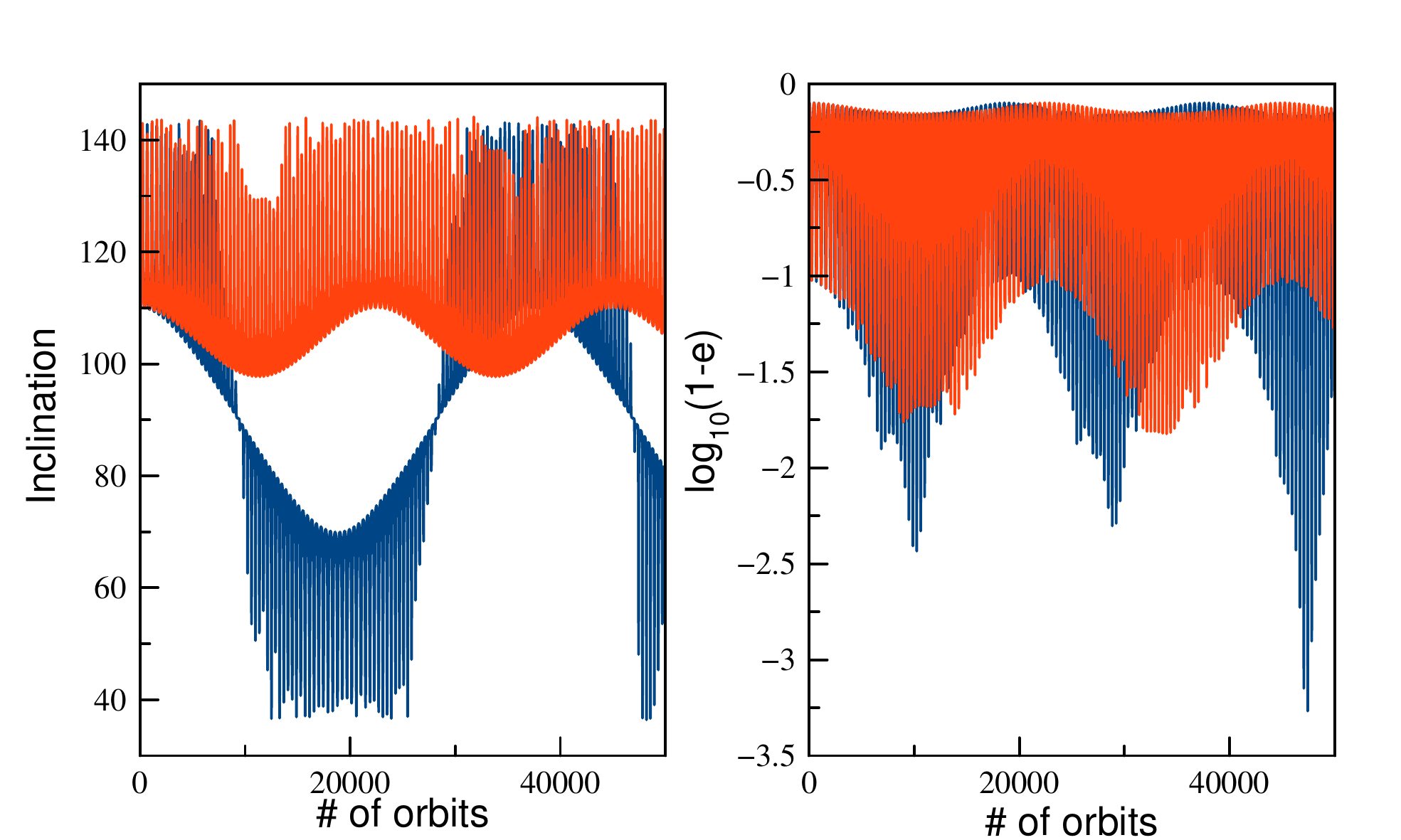}

\caption{Example of the effect of $Q^2$ terms.  A binary consisting of a test body and a solar mass object orbits a distant solar mass object with initial inclination $110$ degrees (a retrograde orbit).  The left panel plots the inclination and the right panel plots $\log_{10}(1-e)$ over 50,000 inner-binary orbits.  The first-order evolution (blue) shows orbital flips and extreme eccentricity excursions.  Including $Q^2$ terms (red) suppresses the flips and the excursions.  The red curves are in good agreement with second-order results and full numerical integrations of \cite{2016MNRAS.458.3060L,2018MNRAS.481.4602L}.  (Color figures in online version.)
\label{fig:LCO} }
\end{center}
\end{figure}

Figure \ref{fig:LCO} shows an example of the effect of these $Q^2$ terms.  This example is displayed here because it has been studied by other 
authors who have recognized the potential importance of second-order perturbations in the orbit evolution equations \cite{2016MNRAS.458.3060L,2018MNRAS.481.4602L,2019MNRAS.490.4756L} (see also \cite{2019MNRAS.487.5630H}).  

In this example, a test particle orbits a $1 \, M_\odot$ body at $1$ astronomical unit (au), perturbed by another $1 \, M_\odot$ body at $10$ au.  Both orbits have initial eccentricities of $0.2$ and an initial relative inclination between the orbital planes of $110$ degrees.  The initial pericenter angles of both orbits are set to zero (see Table \ref{table:params} for a list of the parameters for this and other examples discussed in this paper).   The inclination between the two orbital planes and the inner eccentricity (or $\log_{10} (1-e)$) are plotted over 50,000 inner orbits.  The blue curves are the conventional first-order results, through hexadecapole order, while the red curves include the $Q^2$ terms (the pericenter precessions due to general relativity are negligible in this example).   Without the $Q^2$ terms, the evolution shows clear orbital flips from retrograde to prograde and back, along with excursions to extreme eccentricities ($1-e < 10^{-3}$).   The $Q^2$ terms completely suppress the orbital flips and the extreme eccentricity values.  Figure \ref{fig:LCO} agrees very well with Fig.\ 1 of Luo, Katz and Dong (LKD) \cite{2016MNRAS.458.3060L} and Fig.\ 2 of Lei et al. \cite{2018MNRAS.481.4602L}, which treated the same physical system.  LKD developed an approach called ``Corrected Double Averaging'' (CDA) to go beyond the standard application of the secular approximation at first order in perturbation theory, taking into account the periodic perturbations of the orbit  before averaging over the two orbital timescales.  
The evolution equations resulting from our analysis (Eqs.\ (\ref{eq:QQterms} below) are completely equivalent to those derived by LKD.

These $Q^2$ effects do not suppress all orbital flips.  For low outer-mass systems, such as Hot Jupiters, the $Q^2$ terms have very little effect, as expected.  But for comparable-mass outer bodies, such as in Fig.\ \ref{fig:LCO}, or for high-mass outer bodies, such as a $10 + 90 \, M_\odot$ binary system orbiting a $10^6 \, M_\odot$ massive black hole, the $Q^2$ terms suppress orbital flips.  In other regions of the parameter space, particularly where $Q^2$ terms and octople terms may be comparable we find frequent cases of ``Game of Thrones" style battles for dominance between competing effects, resulting in ragged patterns of minor flips and failed flips, reflecting the sensitivity of three-body dynamics to small effects.

The remainder of this paper presents details.  In Sec.\ \ref{sec:secular}  we show the derivation of the $Q^2$ terms, beginning with the Lagrange planetary equations expressed to quadrupole order, the basics of the two timescale analysis, the special orbit averaging procedure necessitated by second-order perturbation theory, and the final evolution equations.  In Sec.\ \ref{sec:astro} we study the astrophysical implications of the $Q^2$ terms, and in Sec.\ \ref{sec:discussion} we discuss the results.  In Appendix \ref{app:twotimescale} we give a brief review of the two timescale approach, in Appendix \ref{app:secular} we provide the detailed derivation of the averaging procedure applied to second-order terms, and in Appendix \ref{app:Luo} we discuss the equivalence between our results and those of LKD \cite{2016MNRAS.458.3060L}.

\section{Evolution of hierarchical triples to quadrupole-squared order}
\label{sec:secular}

\subsection{Lagrange planetary equations}

\begin{figure}[t]
\begin{center}

\includegraphics[width=3.4in]{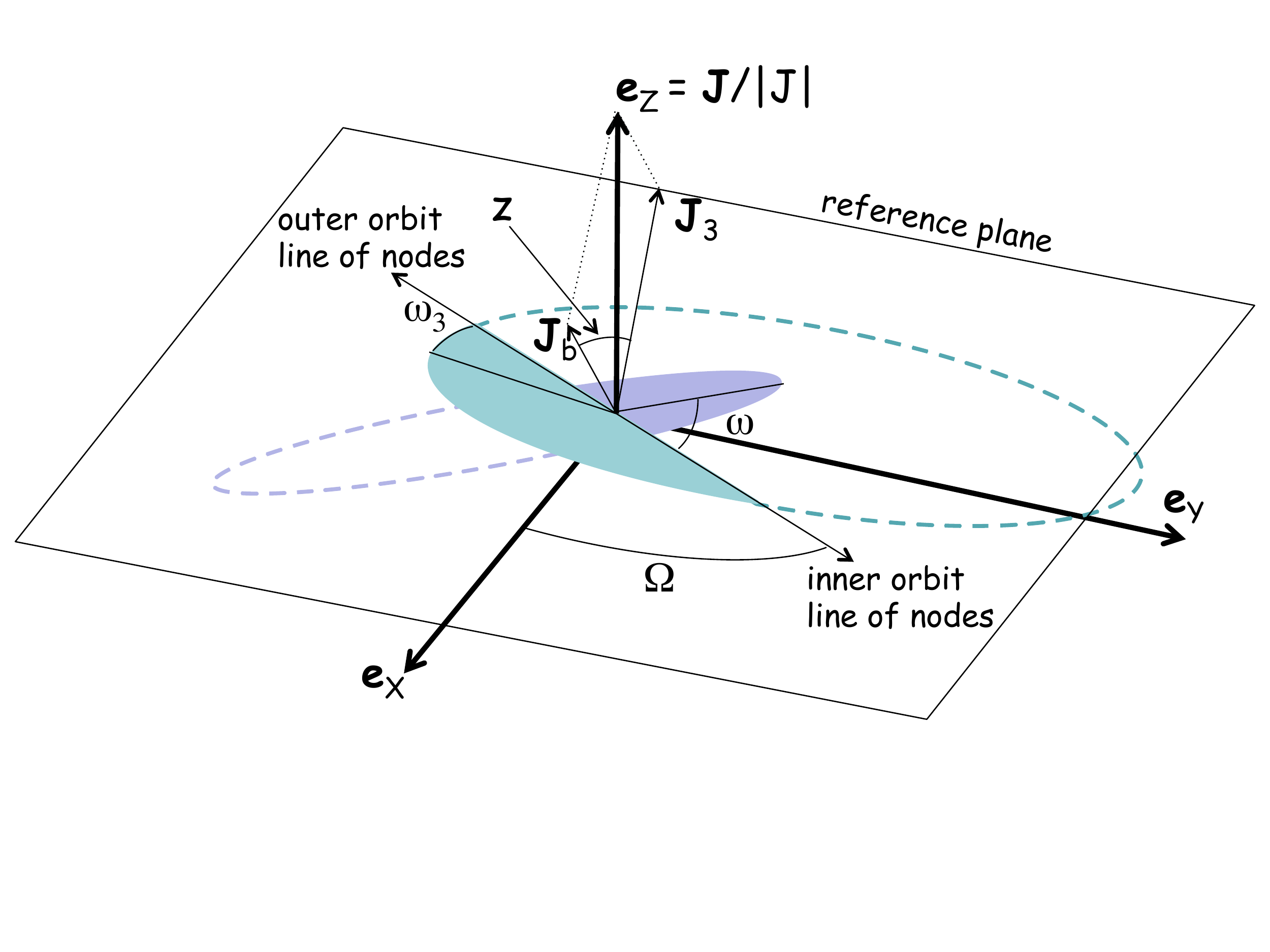}

\caption{Orientation of inner and outer orbits. (Color figures in online version.)
\label{fig:orbits} }
\end{center}
\end{figure}

We consider a hierarchical three-body system illustrated in Fig.\ \ref{fig:orbits}, with bodies 1 and 2 comprising the ``inner'' binary, and with body 3 taken to be the ``outer'' perturbing body.  The orbital separation of the inner binary is assumed to be small compared to that of the outer binary.  We define $m \equiv m_1 + m_2$, $M \equiv m + m_3$, $\eta \equiv m_1m_2/m^2$ with the convention that $m_1 \le m_2$, and $\eta_3 \equiv m_3 m/M^2$ .   To the leading order
in the ratio of $r$ to $R$, where $r \equiv |{\bm x}| = |{\bm x}_1 - {\bm x}_2|$ is the inner binary separation, and $R \equiv |{\bm X}| = |{\bm x}_3 - {\bm x}_{\rm cm}|$ is the separation between the outer body and the center of mass of the inner binary, known as ``quadrupole'' order, the equations of motion take the form
\begin{align}
a^j &= - \frac{Gm n^j}{r^2} + \frac{Gm_3 r}{R^3} \left ( 3 N^j N_n - n^j \right ) 
 \,,
\nonumber \\
A^j &=-  \frac{GM N^j}{R^2} - \frac{3}{2} \frac{GM\eta r^2}{R^4}  \left ( 5 N^j N_n^2 - 2 n^j N_n - N^j  \right )
 \,,
\label{eq2:eom3}
\end{align}
where $\bm{a} \equiv d^2\bx/dt^2$, $\bm{A} \equiv d^2\bm{X}/dt^2$, $\bm{n} \equiv \bx/r$, $\bm{N} \equiv \bm{X}/R$, $N_n \equiv \bm{N} \cdot \bm{n}$, and $G$ is Newton's constant.   

We define the osculating orbit elements of the inner and outer orbits in the standard manner: for the inner orbit, we have the orbit elements $p$, $e$, $\omega$, $\Omega$ and $\iota$, with the definitions
\begin{eqnarray}
r &\equiv& p/(1+e \cos f) \,,
\nonumber \\
{\bm x} &\equiv& r {\bm n} \,,
\nonumber \\
{\bm n} &\equiv& \left [ \cos \Omega \cos(\omega + f) - \cos \iota \sin \Omega \sin (\omega + f) \right ] {\bm e}_X 
\nonumber \\
&&
 + \left [ \sin \Omega \cos (\omega + f) + \cos \iota \cos \Omega \sin(\omega + f) \right ]{\bm e}_Y
\nonumber \\
&&
+ \sin \iota \sin(\omega + f) {\bm e}_Z \,,
\nonumber \\
{\bm \lambda} &\equiv& d{\bm n}/df \,, \quad \hat{\bm h}={\bm n} \times {\bm \lambda} \,,
\nonumber \\
{\bm h} &\equiv& {\bm x} \times {\bm v} \equiv \sqrt{Gmp} \, \bm{\hat{h}} \,,
\label{eq2:keplerorbit1}
\end{eqnarray}
where (${\bm e}_X,\,{\bm e}_Y ,\,{\bm e}_Z$) define a reference basis, with ${\bm e}_Z$ aligned along the total angular momentum of the system, and with the ascending node of the inner orbit oriented at an angle $\Omega$ from the $X$-axis.  From the given definitions, it is evident that ${\bm v} = \dot{r} {\bm n} + (h/r) {\bm \lambda}$ and $\dot{r} = (he/p) \sin f$. 

The outer orbit is defined in the same manner, with orbit elements $P$, $E$, $\omega_3$, $\Omega_3$, and $\iota_3$ replacing $p$, $e$, $\omega$, $\Omega$ and $\iota$, $\bm{\Lambda}$ and $\bm{H}$ replacing $\bm{\lambda}$ and $\bm{h}$, and $F$ replacing $f$.  
The semimajor axes of the two orbits are defined by $a \equiv p/(1-e^2)$ and $A \equiv P/(1-E^2)$.

With the orbits and basis defined this way, it is straightforward to show that
\begin{eqnarray}
\Omega_3 &=& \Omega + \pi \,,
\nonumber \\
J_b \sin \iota &=&  J_3 \sin \iota_3 \,,
\end{eqnarray}
where $J_b =  m \eta \sqrt{Gmp}$ and $J_3 = M\eta_3 \sqrt{GMP}$.
Defining 
\begin{align}
\beta &\equiv \frac{J_b}{J_3} = \frac{\sin \iota_3}{\sin \iota} \,,
\nonumber \\
z &\equiv \iota + \iota_3 \,,
\label{eq2:betaz}
\end{align}
it is straightforward to obtain the relations
\begin{equation}
\cot \iota = \frac{\beta + \cos z}{\sin z} \,,  \qquad \cot \iota_3 = \frac{\beta^{-1} + \cos z}{\sin z} \,,
\label{eq2:inclinations}
\end{equation}
so that only the {\em relative} inclination $z$ between the two orbits is dynamically relevant; given an evolution for $z$ and $\beta$, the individual orbital inclinations can be recovered algebraically from Eqs.\ (\ref{eq2:inclinations}).

From Eqs.\ (\ref{eq2:eom3}), we define the perturbing accelerations $\delta {\bm a} \equiv \bm{a} + Gm{\bm n}/r^2$ and $\delta {\bm A} \equiv \bm{A} + GM{\bm N}/R^2$.
Then, for the inner binary, we define the radial $\cal R$, cross-track $\cal S$ and out-of-plane $\cal W$ components of the perturbing acceleration by
${\cal R} \equiv {\bm n} \cdot \delta {\bm a}$,
 ${\cal S} \equiv {\bm \lambda} \cdot \delta {\bm a}$ and
 ${\cal W} \equiv \bm{\hat{h}} \cdot \delta {\bm a}$,
and we write down the ``Lagrange planetary equations'' for the evolution of the orbit elements,
\begin{eqnarray}
\frac{dp}{dt} &=& 2 \sqrt{\frac{p^3}{Gm}} \frac{{\cal S}}{1+e \cos f} \,,
\nonumber \\
\frac{de}{dt} &=& \sqrt{\frac{p}{Gm}} \left [ \sin f \, {\cal R} + \frac{2\cos f + e +e\cos^2 f}{1+ e\cos f} {\cal S} \right ]\,,
\nonumber \\
\frac{d\varpi}{dt} &=& \frac{1}{e}\sqrt{\frac{p}{Gm}} \left [ -\cos f \, {\cal R} + \frac{2 + e\cos f}{1+ e\cos f}\sin f {\cal S} 
\right ] \,,
\nonumber \\
\frac{d\iota}{dt} &=& \sqrt{\frac{p}{Gm}} \frac{\cos (\omega +f)}{1+ e\cos f} {\cal W} \,,
\nonumber \\
 \frac{d\Omega}{dt} &=& \sqrt{\frac{p}{Gm}} \frac{\sin (\omega +f)}{1+ e\cos f} \frac{\cal W}{\sin \iota}\,.
\label{eq2:lagrange}
\end{eqnarray}
The auxiliary variable $\varpi$ is defined such that the change in pericenter angle is given by $\dot{\omega} = \dot{\varpi} -  \dot{\Omega} \cos \iota$.

For the outer binary, the analogous components of the perturbing acceleration are defined by
${\cal R}_3 \equiv {\bm N} \cdot \delta {\bm A}$,
 ${\cal S}_3 \equiv {\bm \Lambda} \cdot \delta {\bm A}$ and
 ${\cal W}_3 \equiv \bm{\hat{H}} \cdot \delta {\bm A}$.   
The planetary equations for the outer binary take the form of Eqs.\ (\ref{eq2:lagrange}), with suitable replacements of all the relevant variables.

\subsection{Secular evolution of orbit elements to second order}

We now wish to obtain the secular evolution of the orbital elements to second order in the quadrupole perturbation.  This is done using a two-timescale analysis; Appendix \ref{app:twotimescale} gives a brief review of the method. 
Each planetary equation can be written in the generic form
\begin{align}
\frac{dX_\alpha(t)}{dt} &= \varepsilon Q_\alpha (X_\beta(t),t)  \,,
\end{align}
where the $Q_\alpha$ denote the right-hand sides of the Lagrange planetary equations, $\varepsilon$ is a small parameter that characterizes the perturbation.  The solutions will have pieces that vary on a long, secular timescale, of order $1/\varepsilon$ times the orbital timescales, plus periodic pieces that vary on the orbital timescales.  By defining the long-timescale variable $\theta = \varepsilon t$, treating the two variables as independent, and splitting each element into an average part $\tilde{X}_\alpha$ and an ``average-free'' part $Y_\alpha$, 
\begin{equation}
X_\alpha (\theta,t) \equiv \tilde{X}_\alpha (\theta) + \varepsilon Y_\alpha (\tilde{X}_\beta (\theta), t), 
\end{equation}
we can separate each equation into one for the secular evolution of $\tilde{X}_\alpha$ and one for the periodic evolution of $Y_\alpha$, given by
\begin{subequations}
\begin{align}
\frac{d\tilde{X}_\alpha}{d\theta} &= \langle Q_\alpha (\tilde{X}_\beta + \varepsilon Y_\beta, t) \rangle \,,
\label{eq2:aveq}\\
\frac{\partial Y_\alpha}{\partial t} &=\varepsilon {\cal AF} \left (Q_\alpha (\tilde{X}_\beta + \varepsilon Y_\beta, t) \right )  - \varepsilon^2 \frac{\partial Y_\alpha}{\partial \tilde{X}_\gamma} \frac{d\tilde{X}_\gamma}{d\theta} \,,
\label{eq2:avfreeeq}
\end{align}
\label{eq2:maineq}
\end{subequations}
where the average ($\langle \, \rangle$) and average-free ($\cal{AF}$) parts of a function $A$ are defined by
\begin{align}
\langle A \rangle &\equiv \frac{1}{T} \int_0^{T} A(\theta,t) dt \,,
\nonumber \\
 {\cal AF}(A) &\equiv  A(\theta,t) - \langle A \rangle  \,,
\end{align}
where $T$ is a suitable number of periods related to the short-timescale variable $t$.  Working to second order in $\varepsilon$ for the long-timescale evolution, we find (see Appendix \ref{app:twotimescale}) that
\begin{eqnarray}
\frac{d\tilde{X}_\alpha}{dt} &=& \varepsilon \left\langle Q_\alpha^{(0)} \right\rangle 
 + \varepsilon^2 \left\langle {\cal AF} \left (Q_{\alpha,\beta}^{(0)} \right) \int_0^t {\cal AF}\left (Q_{\beta}^{(0)} \right )dt' \right\rangle 
 \nonumber
 \\
&& \quad
+O(\varepsilon^3)\,,
\label{eq2:dXdtfinal}
\end{eqnarray}
where $Q^{(0)}_\alpha \equiv Q_\alpha (\tilde{X}_\beta, t)$, the subscript $,\beta$ denotes $\partial/\partial \tilde{X}_\beta$, and where we have converted from $\theta$ back to $t$. The second-order term contains the $Q^2$ terms in which we are interested.

We now must deal with the fact that the perturbing functions $Q_\alpha$ depend on {\em two} short timescales, the period of the inner orbit and the period of the outer orbit, both small compared to the secular timescale.  It is straightforward to see from the equations of motion that each $Q_\alpha$ is a sum of terms, each of which is product of a function that varies on the inner orbit timescale and depends on the inner orbit elements, and a function that varies on the outer orbit timescale and depends on the outer orbit elements, i.e.
\begin{equation}
Q_\alpha = \sum A_\alpha (X_\beta, t_{\rm in}) M_\alpha  (Z_\beta, t_{\rm out}) \,,
\end{equation}
where the $X_\beta$ and $Z_\beta$ are orbit elements associated with the inner and outer binaries, respectively.   For the leading, $O(\varepsilon)$ term in Eq.\ (\ref{eq2:dXdtfinal}), the average  $\langle Q_\alpha^{(0)} \rangle$ is carried out by adopting the so-called ``secular approximation'', whereby the average of a product of the two functions is the product of their averages, in other words
\begin{equation}
\left\langle Q_\alpha^{(0)} \right\rangle = \sum \left\langle A_\alpha^{(0)} (\tilde{X}_\beta, t_{\rm in}) \right\rangle \left\langle M_\alpha^{(0)}  (\tilde{Z}_\beta, t_{\rm out})  \right\rangle \,,
\label{eq:avQalpha1}
\end{equation}
where
\begin{align}
 \left \langle A_\alpha^{(0)} \right \rangle &\equiv \frac{1}{P_{\rm in}} \int_0^{P_{\rm in}}
 A_\alpha^{(0)} dt \,,
\nonumber \\
\left \langle M_\alpha^{(0)} \right \rangle &\equiv \frac{1}{P_{\rm out}} \int_0^{P_{\rm out}}
 M_\alpha^{(0)} dt \,.
\end{align}
where the two orbital periods are given by $P_{\rm in} = 2\pi \sqrt{a^3/Gm}$ and $P_{\rm out} = 2\pi \sqrt{A^3/GM}$, with the assumption that $P_{\rm in} \ll P_{\rm out}$.  In Appendix \ref{app:secular} we show that this is valid up to corrections of relative order $(P_{\rm in}/P_{\rm out})^2$.  This makes physical sense, because during one inner orbit, the outer body does not move much, so that one can hold it ``fixed'' while averaging over the inner orbit, and then one can average over one outer orbit.  It is important to recognize that the secular approximation ignores the phenomenon of resonances: if the outer orbit is eccentric, then higher harmonics of the basic orbital frequency could be close to the frequency of the inner orbit and generate resonant perturbations, phenomena that are well known theoretically and observationally.   With that caveat in mind, the secular approximation has been a standard tool in studying hierarchical triples (and more complex hierarchical systems).   We will adopt that approximation throughout our work, ignoring resonances completely.

But when we now turn to the $O(\varepsilon^2)$ term in Eq.\ (\ref{eq2:dXdtfinal}), we see that we have a problem because of the integral over the variable $t$.  We need to average the generic quantity ${\cal AF} (AM) \int_0^t {\cal AF}(BN) dt'$, where $A$ and $B$ vary on the short orbital timescale, and $M$ and $N$ vary on the long orbital timescale, and $AM \sim BN \sim Q_\alpha$.  The details are given in Appendix \ref{app:secular}; the result is 
\begin{align}
& \left\langle {\cal AF} (AM) \int_0^t {\cal AF}(BN) dt' \right\rangle 
\nonumber \\
& \qquad =
\left\langle A \right\rangle  \left\langle B \right\rangle  \left\langle {\cal AF} (M) \int_0^t {\cal AF}(N) dt'\right\rangle
\nonumber \\
& \qquad \qquad
+ \left\langle {\cal AF} (A) \int_0^t {\cal AF}(B) dt' \right\rangle  \left\langle MN\right\rangle 
\nonumber \\
& \qquad \qquad
+ O  [P_{\rm in}^2 /P_{\rm out} \times \langle AMBN \rangle ] \,.
\label{eq:secularnew}
\end{align}

It is useful to estimate the sizes of these terms, say for the quadrupole perturbations of Eq.\ (\ref{eq2:eom3}), and for the inner orbit elements.  From Eqs.\ (\ref{eq2:lagrange}), the $Q_\alpha$ are given roughly by
\begin{equation}
Q_\alpha \sim (p/Gm)^{1/2} (Gm_3 r/R^3) \sim P_{\rm in}^{-1} (m_3/m)(a/A)^3 \,,
\end{equation}
(for the semilatus rectum, we use the dimensionless quantity $Q_p /p$) and thus the first-order contribution to $dX_\alpha/dt$ in Eq.\ (\ref{eq2:dXdtfinal}) is just of order $P_{\rm in}^{-1} (m_3/m)(a/A)^3$ (see Eqs.\ (\ref{eq2:quadrupole}) below for explicit formulae).  

Turning to the second-order terms in Eq.\ (\ref{eq2:dXdtfinal}), which have the form of Eq.\ (\ref{eq:secularnew}), we see that the second term in Eq.\ (\ref{eq:secularnew})  is of order
\begin{align}
{\rm term \, 2} &\sim P_{\rm in} \left ( Q_\alpha \right )^2  \nonumber \\
&\sim P_{\rm in}^{-1} \left (\frac{m_3}{m}\right)^2 \left (\frac{a}{A} \right )^6
\end{align}
where the  $P_{\rm in}$ prefactor comes from the integral over the rapidly varying functions.  This term is of the same order in $a/A$ as terms at dotriocontopole order
($P_{\rm in}^{-1} (m_3/m)(a/A)^6$), but has the additional factor of $m_3/m$.  We will ignore the contributions from term 2 henceforth.  
The $O (P_{\rm in}^2/P_{\rm out})$ notation in Eq.\ (\ref{eq:secularnew}) denote terms that are smaller than term 2 by an additional power of $(P_{\rm in}/P_{\rm out})$.   

However, the first term in Eq.\ (\ref{eq:secularnew})  is of order
\begin{align}
{\rm term \, 1} &\sim P_{\rm out} \left ( Q_\alpha \right )^2  
\nonumber \\
&\sim \frac{P_{\rm out}}{P_{\rm in}} \times {\rm term \, 2} 
\nonumber \\
&\sim P_{\rm in}^{-1} \left (\frac{m_3}{m}\right)^2 \frac{1}{(1+m_3/m)^{1/2}} \left (\frac{a}{A} \right )^{9/2} \,,
\end{align}
where the $P_{\rm out}$ prefactor comes from the integral over the slowly varying functions. 
In terms of powers of $a/A$, this contribution lies between octupole-order [$(a/A)^4$] and
hexadecapole-order [$(a/A)^5$] terms.  These quadrupole-quadrupole ($Q^2$) effects arising from ``term 1'' will be the focus of our work.  For low-mass third bodies their effects will be suppressed by the additional factor of $m_3/m$.  However for high-mass third bodies, the effects of these terms could be comparable to or larger than octopole-order perturbations. 

We can manipulate ``term 1'' in Eq.\ (\ref{eq:secularnew}) into a form that can be incorporated into the $O(\varepsilon^2)$ term in (\ref{eq2:dXdtfinal}).  Noting that $AM = Q_{\alpha , \beta}^{(0)}$ and $BN = Q_{\beta}^{(0)}$ in this case, we can write
\begin{align}
&\left\langle A \right\rangle  \left\langle B \right\rangle  \left\langle {\cal AF} (M) \int_0^t {\cal AF}(N) dt' \right\rangle
\nonumber \\
& \qquad
= \left\langle \left [ \langle A \rangle M - \langle A \rangle\langle M  \rangle  \right ]
\int_0^t \left [ \langle B \rangle N - \langle B \rangle\langle N  \rangle  \right ] dt'
\right\rangle
\nonumber \\
& \qquad
= \left\langle \left [ {\rm Av}_1 (Q_{\alpha , \beta}^{(0)}) - \langle Q_{\alpha , \beta}^{(0)}\rangle \right ] 
\right .
\nonumber \\
& \left .
\qquad 
\times \int_0^t  \left [ {\rm Av}_1 (Q_{ \beta}^{(0)}) - \langle Q_{\beta}^{(0)}\rangle \right ] dt'
\right\rangle \,,
\label{eq:epsquaredterm}
\end{align}
where we define
\begin{equation}
 {\rm Av}_1 (Q_{ \beta}^{(0)}) \equiv \frac{1}{P_{\rm in}} \int_0^{P_{\rm in}}
 Q_\beta^{(0)} dt_{\rm in} \,,
\end{equation}
holding $t_{\rm out}$ fixed.

\subsection{Results to $Q^2$ order}

In carrying out the time averages in Eq.\ (\ref{eq2:dXdtfinal}) we must deal with the fact that the functions $Q_\alpha$ do not depend on time explicitly, but instead depend on angular variables that characterize the osculating orbits, such as the ``true anomaly'' $f$ or the ``eccentric anomaly'' $u$.  These are related to time $t$ by the differential equations 
\begin{subequations}
\begin{align}
\frac{df}{dt} &=  \frac{n}{(1-e^2)^{3/2}}  (1+e\cos f )^2 - \frac{d\varpi}{dt} \,,
\label{eq:dfdt}
\\
\frac{du}{dt} &= \frac{n}{1 - e \cos u} - \frac{(1-e \cos u)}{\sqrt{1-e^2}} \frac{d\varpi}{dt} 
\nonumber \\
& \quad - \frac{\sin u}{1-e^2} \frac{de}{dt} \,,
\label{eq:dudt}
\end{align}
\label{eq:dfdtdudt}
\end{subequations}
where $n = (Gm/a^3)^{1/2}$, with analogous formulae for the outer orbit.  The additional terms arise from the fact that $f$ and $u$ are measured from the pericenter, which evolves with time in a complex way (see eg.  \cite{1991ercm.book.....B} for discussion).
The true anomaly $f$ and the eccentric anomaly $u$ are related to each other by \begin{equation}
\sin f = \frac{\sqrt{1-e^2} \sin u}{1- e \cos u} \,, \quad \cos f = \frac{\cos u - e}{1- e \cos u} \,,
\end{equation}
which are compatible with Eqs.\ (\ref{eq:dfdtdudt}).   

These relations between the anomalies and time will also generate $Q^2$ contributions, partly from the additional terms proportional to $d\varpi/dt$ and $de/dt$ in Eqs.\ (\ref{eq:dfdtdudt}), and partly from expanding the orbit elements in these expressions in terms of average and average-free parts.  One way to incorporate these effects is to use Eqs.\ (\ref{eq:dfdtdudt}) to express all the $Q_\alpha$ explicitly in terms of time.  This cannot be done in closed form, but can be done using well-known expressions involving infinite series in powers of $e$ and $E$  \cite{1961mcm..book.....B}.    

An alternative is to use the expressions (\ref{eq:dfdtdudt}) to convert the time-integrals into integrals over $f$ and $F$.   The problem is that we have two orbital angular anomalies ($f$ and $F$) but only one time. 
However, the secular approximation conveniently splits the single time integral into two, one over the inner orbit and one over the outer orbit.  This is true both for the leading term $\langle Q_\alpha^{(0)} \rangle$ as shown in Eq.\ (\ref{eq:avQalpha1}), and for the second-order term, as shown in Eq.\ (\ref{eq:secularnew}).  To carry this out explicitly, we employ the following device:  we define $\widehat{Q}_\alpha$ by
\begin{equation}
\widehat{Q}_\alpha \equiv {Q}_\alpha \frac{dt}{df} \frac{dt}{dF} \,,
\end{equation}
where $dt/df$ and $dt/dF$ are given by the inner and outer orbit versions of Eq.\ (\ref{eq:dfdt}).   Then we express the original $Q_\alpha$ as
\begin{equation}
Q_\alpha \equiv  \widehat{Q}_\alpha \widehat{\frac{df}{dt}}  \widehat{\frac{dF}{dt}} \,,
\end{equation}
where
the quantities $\widehat{df/dt}$ and  $\widehat{dF/dt}$ are meant to be abstract placeholders for the ultimate conversion of $dt$ into either $df$ or $dF$ {\em after} applying the secular approximation.  Returning to our two-timescale analysis, we make the expansion
\begin{equation}
Q_\alpha = \left ( \widehat{Q}_\alpha^{(0)} +\varepsilon \widehat{Q}_{\alpha , \beta}^{(0)} Y_\beta \right ) \widehat{\frac{df}{dt}}  \widehat{\frac{dF}{dt}} \,,
\end{equation}
so that the derivatives with respect to the orbit elements in $ \widehat{Q}_{\alpha , \beta}^{(0)}$ now automatically include their contributions to $dt/df$ and $dt/dF$.   With this trick, the $O(\varepsilon)$ term, Eq.\ (\ref{eq:avQalpha1}) becomes
\begin{align}
\left\langle Q_\alpha^{(0)} \right\rangle &= \sum \frac{n}{2\pi} \int_0^{2\pi} \widehat{A}_\alpha df  \frac{N}{2\pi} \int_0^{2\pi} \widehat{M}_\alpha dF 
\nonumber \\
&= \frac{n}{2\pi} \int_0^{2\pi} \frac{N}{2\pi} \int_0^{2\pi} \widehat{Q}_\alpha df   dF \,. 
\end{align}
In a similar way, we can write the $O(\varepsilon^2)$ expression (\ref{eq:epsquaredterm}) the explicit form
\begin{align}
&\left\langle {\cal AF} \left (Q_{\alpha,\beta}^{(0)} \right) \int_0^t {\cal AF}\left (Q_{\beta}^{(0)} \right )dt' \right\rangle 
\nonumber \\
&\to  \frac{N}{2\pi} \int_0^{2\pi}  \left \{ \left [\frac{n}{2\pi} \int_0^{2\pi}  \widehat{Q}_{\alpha , \beta}^{(0)} df - \left \langle \widehat{Q}_{\alpha , \beta}^{(0)} \right \rangle \widehat{\frac{dt}{dF}} \right ]
\right.
\nonumber \\ 
\nonumber \\ 
& 
\left .\quad \times \int_0^F \left [\frac{n}{2\pi} \int_0^{2\pi}  \widehat{Q}_{ \beta}^{(0)} df - \left \langle \widehat{Q}_{ \beta}^{(0)} \right \rangle \widehat{\frac{dt}{dF'}} \right ] dF' \right \} dF \,,
\end{align}
where here $\widehat{dt/dF} =\tilde{P}^{3/2}M^{-1/2} (1+\tilde{E} \cos F)^{-2}$.

It is simple to show that, in the case of $Q^2$ terms, the correction terms in Eqs.\ (\ref{eq:dfdtdudt}) involving $d\varpi/dt$ and $de/dt$ generate effects at the same level as ``term 2'' in Eq. (\ref{eq:secularnew}), in other words of dotriocontupole order with an extra $m_3/m$ factor.   Only the periodic variations of the orbit elements within the leading terms in Eqs.\ (\ref{eq:dfdtdudt}) generate contributions of interest.  For practical reasons, we use the eccentric anomaly $u$ for the inner orbit variables and the true anomaly $F$ for the outer orbit variables.  The algebraic work is carried out using {\em Maple}.

After carrying out the orbital averages, we convert from time $t$ to a dimensionless time scaled by the inner orbital period, namely 
\begin{equation}
\tau \equiv \frac{t}{P_{\rm in}} = \frac{t}{2\pi} \left( \frac{Gm}{a^3} \right )^{1/2} \,.
\end{equation}
With this scaling, the entire secular dynamics depends on the three dimensionless parameters:
\begin{equation}
\alpha \equiv \frac{m_3}{m} \,, \quad \eta \equiv \frac{m_1m_2}{m^2} \,,  \quad \epsilon \equiv \frac{a}{A}   \,.
\end{equation}
In terms of these parameters, the quantity $\beta = J_b/J_3$ is given by
\begin{equation}
\beta = \eta \frac{(1+\alpha)^{1/2}}{\alpha}  \epsilon^{1/2} \left ( \frac{1-e^2}{1-E^2} \right )^{1/2} \,.
\end{equation}

At quadrupole order, we obtain the standard eccentric Kozai-Lidov results: 
\begin{widetext}
\begin{align}
\frac{de}{d\tau} & = \frac{15 \pi}{2} \alpha \epsilon^3  \frac{e(1-e^2)^{1/2}}{(1-E^2)^{3/2}} \sin^2 z \sin \omega \cos \omega \,,
\nonumber \\
\frac{d \iota}{d\tau} &= - \frac{15 \pi}{2} \alpha \epsilon^3  \frac{e^2}{(1-e^2)^{1/2}(1-E^2)^{3/2}} \sin z \cos z  \sin \omega \cos \omega \,,
\nonumber \\
\frac{d\Omega}{d\tau} &= - \frac{3\pi}{2}  \alpha \epsilon^3 \frac{1}{(1-e^2)^{1/2}(1-E^2)^{3/2}} \frac{\sin z \cos z}{\sin \iota} \left (1+4e^2 - 5e^2 \cos^2 \omega \right ) \,,
\nonumber \\
\frac{d\varpi}{d\tau} &= \frac{3\pi}{2}  \alpha \epsilon^3 \frac{(1-e^2)^{1/2}}{(1-E^2)^{3/2}} \left [ 1- \sin^2 z \left ( 4 - 5 \cos^2 \omega \right ) \right ] \,,
\nonumber \\
\frac{dE}{d\tau} & = 0 \,,
\nonumber \\
\frac{d\iota_3}{d\tau} &= - \frac{15 \pi}{2} \eta (1+\alpha)^{1/2}  \epsilon^{7/2}  \frac{e^2}{(1-E^2)^{2}}  \sin z  \sin \omega \cos \omega \,,
\nonumber \\
\frac{d\varpi_3}{d\tau} &= \frac{3\pi}{4}   \eta (1+\alpha)^{1/2}  \epsilon^{7/2} \frac{1}{(1-E^2)^{2}} \left [ 2+3e^2 - 3\sin^2 z \left (1+4e^2 -5e^2 \cos^2 \omega \right ) \right ] \,.
\label{eq2:quadrupole}
\end{align}
From quadrupole through hexadecapole order (and probably to all orders), it is well-known that $p$, $e$, $P$, and $E$ evolve in such a way that the semimajor axes $a$ and $A$ are constant, in other words
\begin{equation}
\frac{da}{d\tau} = \frac{dA}{d\tau} = 0 \,.
\end{equation}
The $Q^2$ contributions for the inner orbit also yield this result.  The remaining $Q^2$ equations for the inner orbit elements are given by
\begin{align}
\frac{de}{d\tau} & = \frac{15\pi}{32} \frac{\alpha^2 \epsilon^{9/2}}{(1+\alpha)^{1/2}}  \frac{e(1-e^2)}{(1-E^2)^3} \biggl [  3 (3+2E^2) \cos z \sin^2 z \sin 2\omega
\nonumber \\
& \qquad
-\frac{5}{2} E^2 H(E) \left ( (1+\cos z)^2 (2-3\cos z)\sin(2\omega-2\omega_3) -  (1-\cos z)^2 (2+3\cos z)\sin(2\omega+2\omega_3) \right ) \biggr ]\,,
\nonumber \\
\frac{d \iota}{d\tau} &= - \frac{15\pi}{32} \frac{\alpha^2 \epsilon^{9/2}}{(1+\alpha)^{1/2}}  
\frac{\sin z}{(1-E^2)^3} \biggl [ 3e^2 (3+2E^2) \cos^2 z  \sin 2\omega
\nonumber \\
& \qquad
+\frac{1}{2} E^2 H(E) \biggl ( 5e^2(1+\cos z) (2-3\cos z)\sin(2\omega-2\omega_3) +  5e^2(1-\cos z) (2+3\cos z)\sin(2\omega+2\omega_3) 
\nonumber \\
& \qquad
-2(2-17e^2) \cos(z) \sin (2\omega_3) \biggr ) \biggr ]\,,
\nonumber \\
\frac{d\Omega}{d\tau} &= - \frac{3\pi}{64} \frac{\alpha^2 \epsilon^{9/2}}{(1+\alpha)^{1/2}} 
\frac{1}{(1-E^2)^3} \frac{\sin(z)}{\sin(i)} \biggl [ (3+2E^2) \biggl ( 2+33e^2-3(2-17e^2)\cos^2 z +15e^2 (1-3\cos^2 z) \cos 2\omega \biggr )
\nonumber \\
& \qquad
-\frac{5}{2} E^2 H(E) \biggl ( 5e^2(1+\cos z) (1-9\cos z)\cos(2\omega-2\omega_3) +  5e^2(1-\cos z) (1+9\cos z)\cos(2\omega+2\omega_3) 
\nonumber \\
& \qquad
+2(2-17e^2) (1-3\cos^2 z) \cos 2\omega_3 \biggr ) \biggr ]\,,
\nonumber \\
\frac{d\varpi}{d\tau} &=   \frac{3\pi}{64} \frac{\alpha^2 \epsilon^{9/2}}{(1+\alpha)^{1/2}} 
\frac{1}{(1-E^2)^3} \biggl [ (3+2E^2) \cos z \biggl ( 64-99e^2 + 3(12-17e^2)\cos^2 z +15(2-3e^2) \sin^2 z \cos 2\omega \biggr )
\nonumber \\
& \qquad
-\frac{5}{2} E^2 H(E) \biggl ( 5(2-3e^2) \bigl [(1+\cos z)^2 (2-3\cos z)\cos(2\omega-2\omega_3) -  (1-\cos z)^2 (2+3\cos z)\cos(2\omega+2\omega_3) \bigr ]
\nonumber \\
& \qquad
-6(12-17e^2) \sin^2 z \cos z \cos 2\omega_3 \biggr ) \biggr ]\,,
\label{eq:QQterms}
\end{align}
\end{widetext}
where 
\begin{equation}
H(E) = 1 - \frac{2(1-E^2)}{5(1+\sqrt{1-E^2})^2} \,,
\end{equation}
with $H(0) = 0.9$ and $H(1)=1$.  

We note that in the limit $E=0$ and $\eta=0$, both the quadrupole terms and the $Q^2$ terms satisfy the Kozai-Lidov property that 
$\cos \iota \sqrt{1-e^2} = {\rm constant}$, embodying the approximate conservation of the inner orbit's angular momentum component $L_z$ perpendicular to the plane of the outer orbit, reflecting the fact that in this limit,  the outer mass can be averaged into an axially symmetric ``wire''.   We also note that, since the original equations of motion (\ref{eq2:eom3}) are invariant under time reversal $t \to -t$, so too should the secular evolution equations.  Because the orbit elements $\iota$, $\Omega$ and $\omega$ are defined by components of the angular momentum vector $\bm h$ and the Runge-Lenz vector ${\bm A} = {\bm v} \times {\bm h}/Gm - {\bm n}$, which satisfy ${\bm h} \to - {\bm h}$ and ${\bm A} \to {\bm A}$, then the elements behave according to $\iota \to \pi - \iota$, $\Omega \to \pi + \Omega$ and $\omega \to \pi - \omega$.  The elements $e$ and $p$ do not change under $t \to -t$.  The orbit elements for the outer orbit transform in the same way; note that $z = \iota + \iota_3$ transforms as $z \to 2\pi - z$.  It is then straightforward to show that the quadrupole and $Q^2$ evolution equations are invariant under time reversal.

Finally we remark that, in obtaining the $Q^2$ equations, we only included derivatives with respect to the inner orbit elements in the $O(\varepsilon^2)$ term in Eq.\ (\ref{eq2:dXdtfinal}).  Equation (\ref{eq2:quadrupole}) shows that perturbations of the outer orbit elements are a factor $\sim \eta (1+\alpha)^{1/2} \alpha^{-1} \epsilon^{1/2}$ relative to perturbations of the inner orbit elements.  As we will see in the next subsection, $Q^2$ effects are important, i.e. larger than octupole or hexadecapole effects {\em only} when $\alpha \gg 1$, in which case the terms we have neglected are small corrections.

\section{Astrophysical implications}
\label{sec:astro}

Using well-known results for the conventional quadrupole, octupole and hexadecapole perturbations (eg.\ \cite{2017PhRvD..96b3017W}), we can estimate the timescales for these perturbations.  In units of the inner orbit period, they are
\begin{align}
T_{\rm Quad} &\sim \frac{(1-E^2)^{3/2}}{\alpha \epsilon^3} \,,
\nonumber \\
T_{\rm Oct} &\sim \frac{(1-E^2)^{5/2}}{E \Delta \alpha \epsilon^4} \,,
\nonumber \\
T_{\rm Hex} &\sim \frac{(1-E^2)^{7/2}}{(1-3\eta) \alpha \epsilon^5} \,,
\nonumber \\
T_{\rm Q^2} &\sim \frac{\sqrt{1+\alpha}}{\alpha^2 }\frac{(1-E^2)^{3}}{\epsilon^{9/2}} \,,
\end{align}
where $\Delta = (m_2-m_1)/m = \sqrt{1-4\eta}$ (recall that $0 \le \eta \le 1/4$).

\begin{figure}[t]
\begin{center}

\includegraphics[width=3.4in]{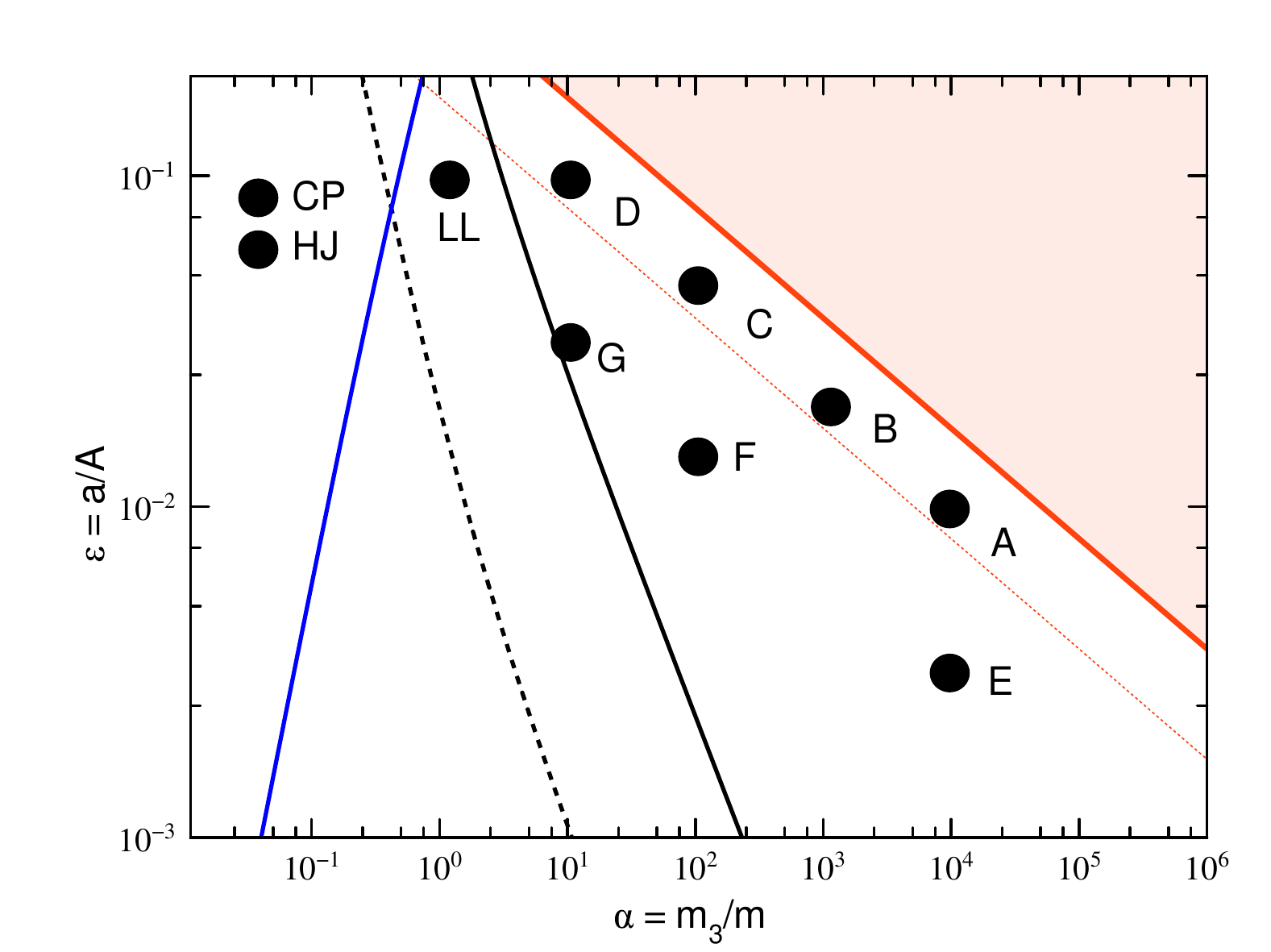}

\caption{Parameter space (Color figures in online version.)
\label{fig:parameterspace} }
\end{center}
\end{figure}

Figure \ref{fig:parameterspace} displays a number of curves that delineate the parameter space that is relevant for the $Q^2$ effects.  The red curve denotes where the conventional quadrupole, or Kozai-Lidov timescale $T_{\rm Quad}$ is around 10 inner or 7 outer orbital periods (for $E =0.6$).  Above this curve, the accuracy of any results based on small perturbations, the hierarchical assumption and the secular approximation is questionable.  We treat this pink area as ``forbidden'' or non-perturbative.   The dotted red curve denotes a KL timescale of around 100 inner, or 65 outer  orbits.   The black curves denote where the octupole timescale equals the $Q^2$ timescale, assuming $\Delta \approx 1$ or $\eta \ll1$.  The solid curve is for $E = 0.6$, while the dotted curve is for $E=0.1$ (when $E=0$, or when $\eta = 1/4$, octupole-order effects on the inner orbit vanish).  To the right of this curve, $T_{\rm Q^2} < T_{\rm Oct}$, in other words, $Q^2$ effects may dominate octupole effects.  Not surprisingly, this is the high-mass regime for the third body.   The blue curve is where $T_{\rm Q^2} = T_{\rm Hex}$; to the right of this curve but to the left of the black curves $Q^2$
 effects may dominate hexadecapole effects but not octupole effects.  
 
So far we have restricted our attention to Newtonian gravity.   In the real world, general relativity (GR) should be included, and indeed it is well known that the simplest quadrupole-order Kozai-Lidov oscillations can be strongly suppressed if the rate of relativistic advance of the pericenter of the inner binary is large enough \cite{1997Natur.386..254H}.   Including the leading contribution of general relativity forces us to introduce an additional dimensionless parameter $\delta$ to the problem, given by
\begin{equation}
\delta \equiv \frac{Gm}{c^2 a} = 9.8705 \times 10^{-9} \left (\frac{m}{M_\odot} \right ) \left (\frac{{\rm au}}{a} \right )\,,
\end{equation}
where $c$ is the speed of light.   The dominant effect is to add to the pericenter advances of the two orbits the terms
\begin{align}
\frac{d\varpi}{d\tau} &= 6\pi \frac{\delta}{1-e^2} \,,
\nonumber\\
\frac{d\varpi_3}{d\tau} &= 6\pi \frac{\delta(1+\alpha)^{3/2} \epsilon^{5/2}}{1-E^2} \,,
\end{align}  
where both are expressed in terms of time scaled by the inner orbital period.  These GR precessions will be included in all our numerical evolutions.

In all the examples to be presented, blue curves correspond to turning the $Q^2$ terms  off and red curves correspond turning the $Q^2$ terms on.  In all cases, the linear in $\varepsilon$ contributions from quadrupole through hexadecapole order are included (see \cite{2017PhRvD..96b3017W} for the full set of equations used.)

\begin{figure}[t]
\begin{center}

\includegraphics[width=3.6in]{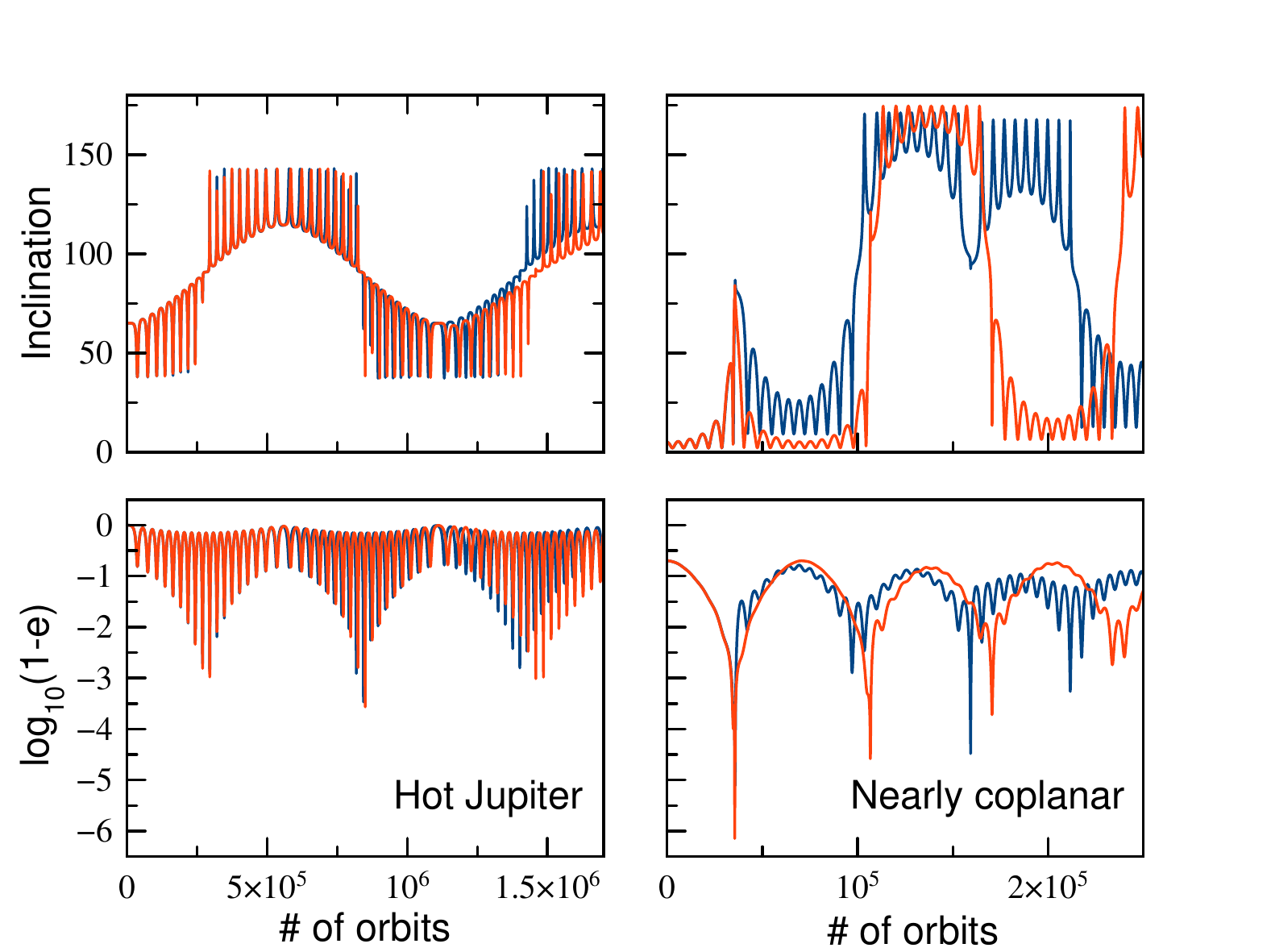}

\caption{Left: Orbital flips and eccentricity excursions in a Jupiter-Sun system perturbed by a distant brown dwarf.  Right:  Orbital flips from nearly coplanar orbits in a similar Jupiter-Sun-brown dwarf system.    Blue: Quadrupole through hexadecapole order, including GR.  Red: $Q^2$ terms added. Parameters and initial orbit elements are listed in Table \ref{table:params}.  (Color figures in online version. )
\label{fig:HJ-CP} }
\end{center}
\end{figure}

\begin{center}
\begin{table*}[t]
\caption{Physical parameters and initial conditions for selected case studies} 
\begin{tabular}{l@{\hskip 0.5cm}c@{\hskip 0.5cm}c@{\hskip 0.5cm}c@{\hskip 0.5cm}c@{\hskip 0.5cm}c@{\hskip 0.5cm}c@{\hskip 0.5cm}c@{\hskip 0.5cm}c@{\hskip 0.5cm}c@{\hskip 0.5cm}c}
\hline
Case&$m_1$&$m_2$&$m_3$&$a$ (au)&$A (au)$&$e$&$E$&$z$&$\omega$&$\omega_3$ \\
\hline 
LL&$0$&$M_\odot$&$M_\odot$&1&10&0.02&0.2&110&0&0\\
HJ&$M_J$&$M_\odot$&$40\, M_J$&6&100&0.001&0.6&65&45&0\\
CP&$M_J$&$M_\odot$&$0.03\, M_\odot$&4&50&0.8&0.6&5&0&0\\
A&$M_\odot$&$100\, M_\odot$&$10^6 \, M_\odot$&$0.01$&$1$&$0.01$&$0.6$&$85$&$0$&$0$\\
B&$M_\odot$&$100\, M_\odot$&$10^5 \, M_\odot$&$0.01$&$0.46$&$0.01$&$0.6$&$85$&$0$&$0$\\
C&$M_\odot$&$100\, M_\odot$&$10^4 \, M_\odot$&$0.01$&$0.21$&$0.01$&$0.6$&$85$&$0$&$0$\\
D&$M_\odot$&$100\, M_\odot$&$10^3 \, M_\odot$&$0.01$&$0.1$&$0.01$&$0.6$&$85$&$0$&$0$\\
E&$M_\odot$&$100\, M_\odot$&$10^6 \, M_\odot$&$1$&$315$&$0.01$&$0.6$&$85$&$0$&$0$\\
F&$M_\odot$&$100\, M_\odot$&$10^4 \, M_\odot$&$1$&$68$&$0.01$&$0.6$&$85$&$0$&$0$\\
G&$M_\odot$&$100\, M_\odot$&$10^3 \, M_\odot$&$1$&$32$&$0.01$&$0.6$&$85$&$0$&$0$\\
\hline
\end{tabular}
\label{table:params}
\end{table*}
\end{center}

As we have discussed, when $m_3 \ll m$, we expect $Q^2$ effects to be suppressed relative to octupole and hexadecapole effects, because of the extra factor of $m_3/m$.  This is borne out by two specific examples, denoted ``HJ'' (hot Jupiters) and ``CP'' (coplanar flips) in Fig.\ \ref{fig:parameterspace}.  

In the hot Jupiter example \cite{2011Natur.473..187N,2013MNRAS.431.2155N},  the inner binary is a Jupiter-mass planet orbiting a solar-mass star with $a=6$  au, perturbed by a brown-dwarf star with a mass of $40 M_J$ and $A = 100$ au (see Table \ref{table:params} for a list of parameters for all the cases examined).    With $M_\odot = 1047 M_J$, the parameters (including the GR parameter)  take the values
\begin{align}
\alpha &= 0.0382 \,, \quad \epsilon = 0.06 \,, 
\nonumber \\
 \eta &= 9.53 \times 10^{-4}  \,, \quad
\delta = 1.65 \times 10^{-9} \,.
\end{align}
The initial conditions chosen in \cite{2011Natur.473..187N,2013MNRAS.431.2155N} were
\begin{equation}
e =  0.001 \,, \, E = 0.6 \,, \, z = 65^{\rm o} \,, \, \omega = 45^{\rm o} \,, \, \omega_3 = 0^{\rm o} \,.
\end{equation}
We evolve the secular planetary equations for $1.7 \times 10^6$ orbits of the inner binary (corresponding to about $2.5 \times 10^7$ years).   The results are plotted in the left panel of Fig.\ \ref{fig:HJ-CP}, with blue and red denoting evolutions without and with $Q^2$ terms, respectively.  Initially the system undergoes Kozai-Lidov type oscillations in inclination $z$ but with the maximum value of $z$ rising steadily; when $z$ reaches $90^{\rm o}$, the orbit becomes retrograde and the oscillations ``flip''.  Later the orbit flips back to prograde, and so on.  Meanwhile, the eccentricity migrates to very large values in the vicinity of each orbital flip.  The $Q^2$ terms make very little difference in this case.     

A second example in the low-$m_3$ regime is the so-called ``nearly coplanar flips'' (CP) case
\cite{2014ApJ...785..116L}.  
The inner system is again a Jupiter-Sun binary with $a=4$ au, perturbed by a brown dwarf, with $m_3 = 0.03 \, M_\odot$ and $A = 50$ au.   The parameters then have the values
 \begin{align}
\alpha &= 0.030 \,, \quad \epsilon = 0.08 \,, 
\nonumber \\
 \eta &= 9.53 \times 10^{-4} \,, \quad \delta = 2.47 \times 10^{-9} \,,
\end{align}
and the initial conditions are
\begin{equation}
e =  0.8 \,, \, E = 0.6 \,, \, z = 5^{\rm o} \,, \, \omega = 0^{\rm o} \,, \, \omega_3 = 0^{\rm o} \,.
\end{equation}

We evolve the equations for $2.5 \times 10^5$ inner orbits ($2 \times 10^6$ years),     The results are shown in the right panel of Fig.\ \ref{fig:HJ-CP}.   The qualitative behavior consisting of orbital flips and excursions to large eccentricity is the same whether the $Q^2$ terms are on or off; only the fine details are different (a flip aborted in one case, but achieved in the other), reflecting the strong sensitivity of three-body evolutions to small changes in the dynamics.  

 We now turn to the high-outer-mass regime, where $Q^2$ effects might be more important.   We first consider a sequence of examples labeled A through D, lying along the dotted red line in Fig.\ \ref{fig:parameterspace}.  For $E=0.6$, this line corresponds to $T_{\rm Quad} \sim 100$ inner orbital periods.    But it also corresponds to  $T_{Q^2} \sim 700$ periods.  It is worth noting that, along this curve, $P_{\rm out}/P_{\rm in} \simeq (\alpha \epsilon^3)^{1/2} \sim 15$.

\begin{figure}[t]
\begin{center}

\includegraphics[width=3.6in]{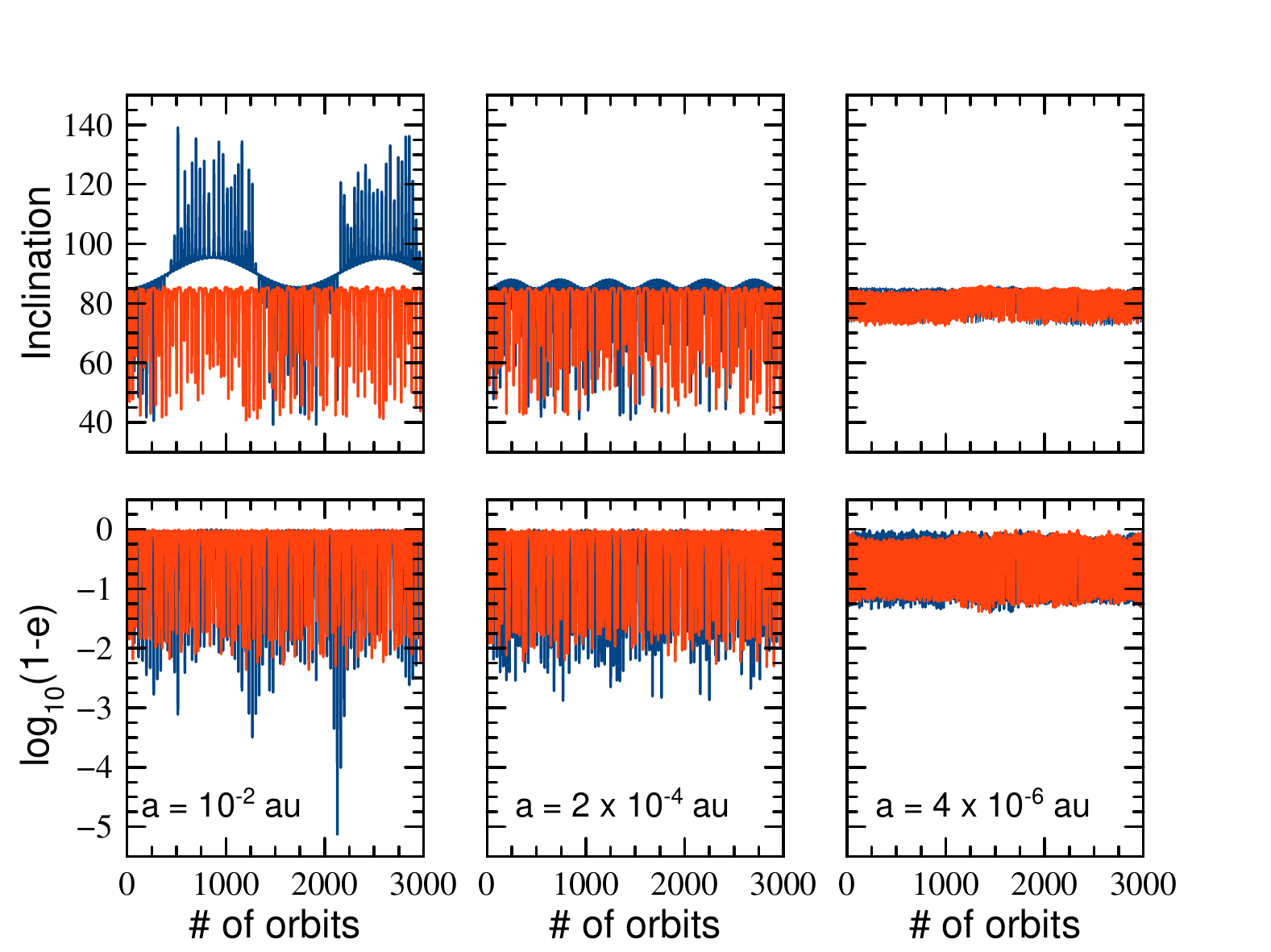}

\caption{Case A. $1 \,M_\odot + 100 \, M_\odot$ binary orbiting a $10^6 \,M_\odot$ black hole.  Left: with GR precessions unimportant, linear multipole terms generate orbital flips and extreme eccentricities (blue), but $Q^2$ terms suppress these effects (red).  Middle: modest GR precessions suppress extreme effects. Right: strong GR precessions suppress even Kozai-Lidov oscillations  (Color figures in online version.)
\label{fig:caseA} }
\end{center}
\end{figure}

\begin{figure}[t]
\begin{center}

\includegraphics[width=3.6in]{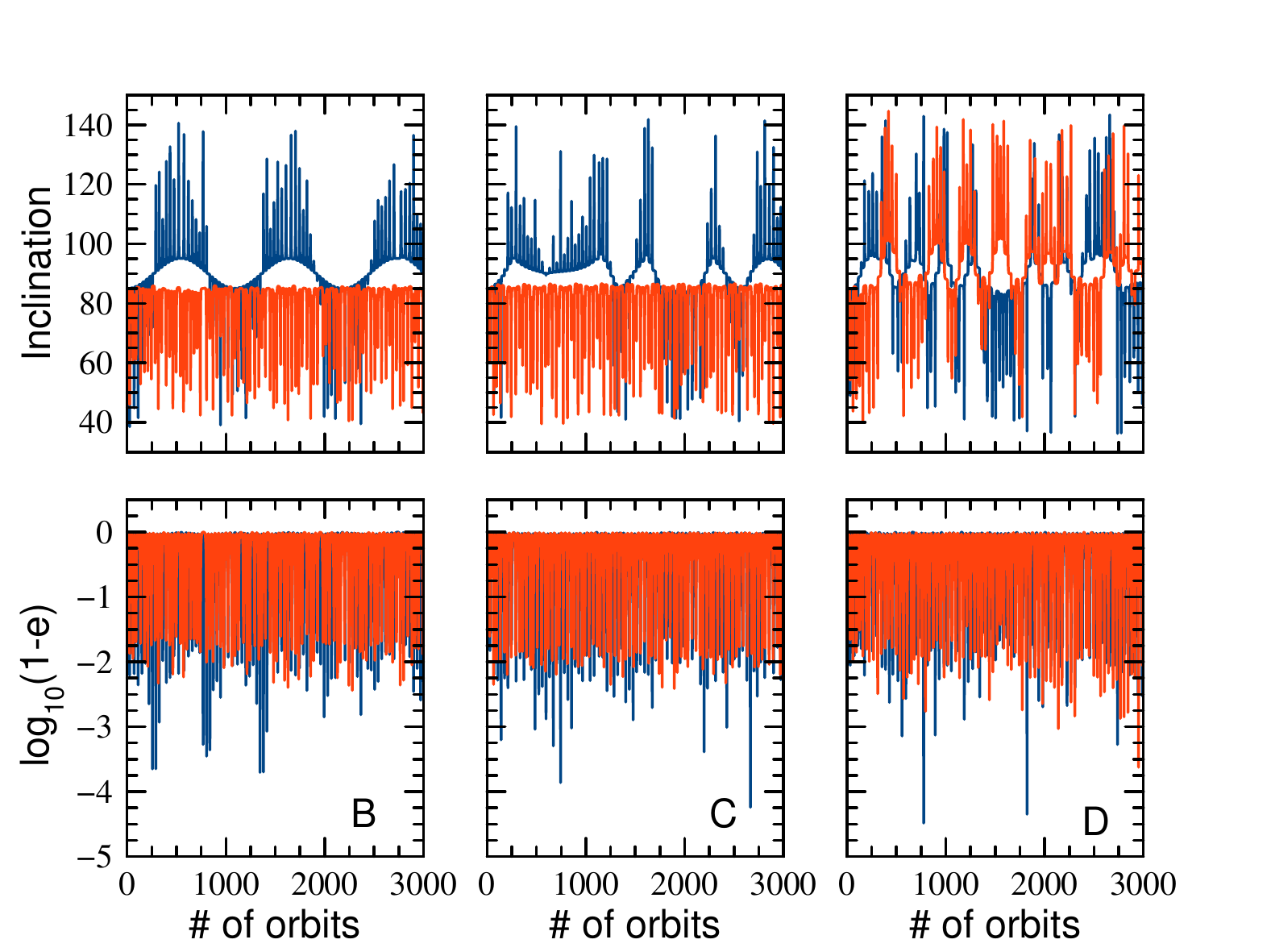}

\caption{Cases B, C \& D.  Left and middle:  For $m_3/m = 10^3$ and $10^2$, $Q^2$ terms suppress orbital flips.  Right: For $m_3/m = 10$, $Q^2$ and octupole terms are comparable in size, leading to complex patterns of orbital flips. (Color figures in online version.)
\label{fig:caseBCD} }
\end{center}
\end{figure}

 Case A is characterized by $\alpha = 10^4$ and $\epsilon = 10^{-2}$, with $\eta = 0.01$.  A specific realization would be a solar mass neutron star and a $100 \, M_\odot$ black hole orbiting a $10^6 \, M_\odot$ black hole.  Holding these parameters fixed, we vary the semimajor axis of the inner orbit from $10^{-2}$ au to $2 \times 10^{-4}$ au to $4 \times 10^{-6}$ au as a way of dialing up the precession effects of GR while keeping the basic Newtonian dynamics unchanged.   Figure \ref{fig:caseA} shows the results, for evolutions over 3000 inner orbits. In the left panel, where the timescale for GR precessions is over 50,000 inner orbits, i.e. where GR has negligible impact, the conventional first-order contributions  produce orbital flips and excursions to extreme eccentricities (blue).  Turning on the $Q^2$ terms (red) completely suppresses the flips and the most extreme eccentricities, although values of $e \sim 0.99$ may still be reached.   Turning on GR precessions (middle panel) suppresses flips and extreme eccentricities whether the $Q^2$ terms are turned on or off.  Even stronger GR effects (right panel), with a timescale of only $\sim  21$ inner orbits, produce the well-known suppression of the amplitude of Kozai-Lidov oscillations.  Here the $Q^2$ terms make very little difference.
 
Cases B, C and D have the same value of $\alpha \epsilon^3$ as Case A, hence the same approximate ratio of quadrupole to $Q^2$ amplitudes, but $\alpha = 10^3 , 10^2$ and $10$, respectively.   For a $1 \, M_\odot + 100 \, M_\odot$ inner binary, we choose $a = 10^{-2}$ au, so that GR effects are negligible (the precession timescale is $\sim 5 \times 10^5$ inner orbital periods).  For cases B and C, the left and middle panels of Fig.\ \ref{fig:caseBCD} show the same pattern of orbital flips being suppressed by the $Q^2$ terms.   However, case D shows a complex array of orbital flips whether the $Q^2$ terms are on or off.  Case D is in the region of parameter space (Fig.\ \ref{fig:parameterspace}) where the timescale for octupole terms is becoming comparable to that of the  $Q^2$ terms, and that panel of Fig.\ \ref{fig:caseBCD} indicates the pitched battle for supremacy between the two kinds of effects.

\begin{figure}[t]
\begin{center}

\includegraphics[width=3.6in]{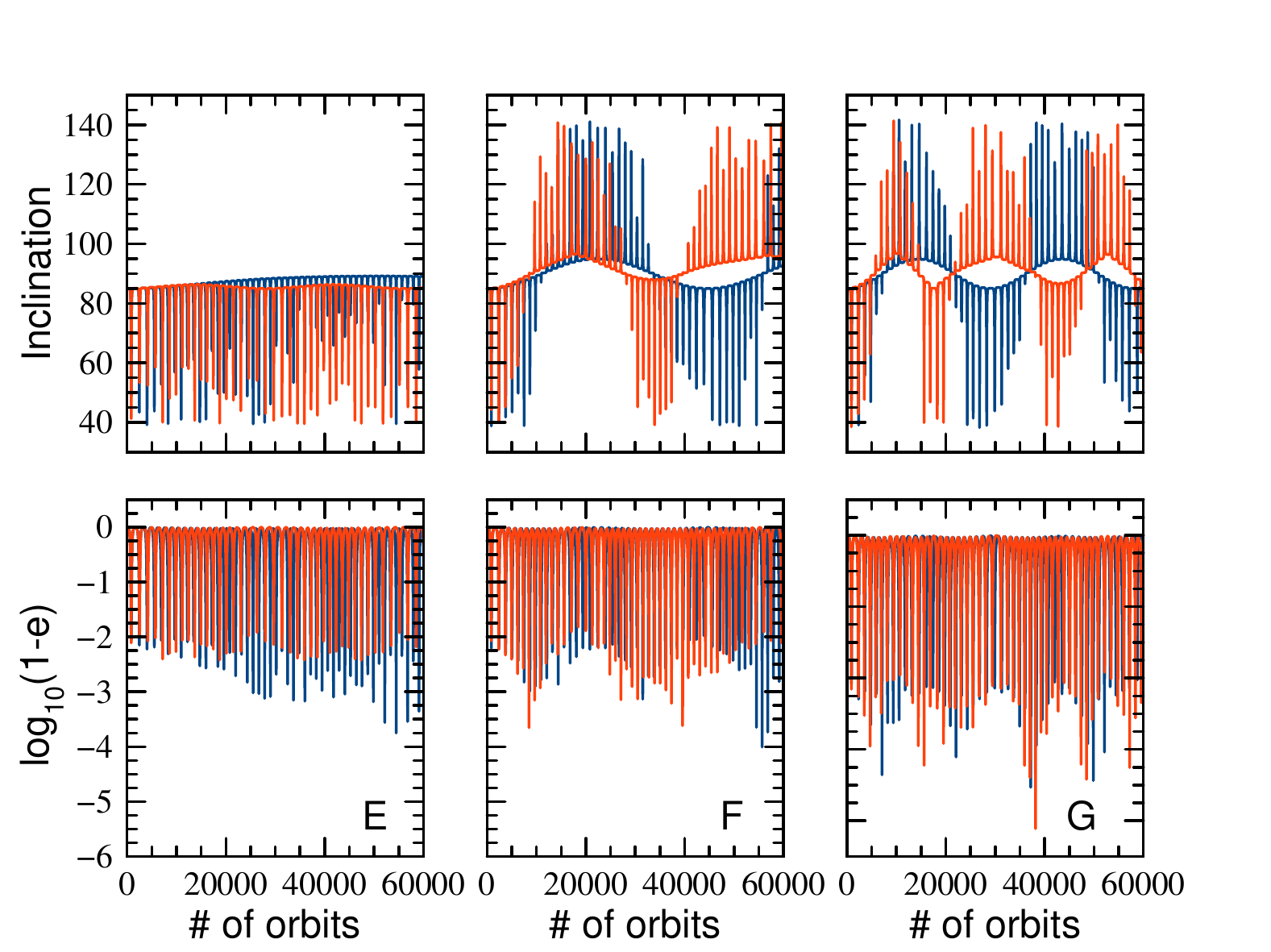}

\caption{Cases E, F \& G.  Left:  For $m_3/m = 10^4$, there are no flips with or without $Q^2$ terms. Middle and Right: For $m_3/m = 100$ and $m_3/m = 1.25$, $Q^2$ and octupole terms are closer in size, leading to complex patterns of orbital flips. (Color figures in online version.)
\label{fig:caseEFG} }
\end{center}
\end{figure}

Cases E, F, and G show a similar pattern.   The $Q^2$ terms are now  weak compared to quadrupole terms, and are becoming comparable to octupole terms.  For Case E, $\epsilon$ is very small, and thus the octupole terms are themselves small compared to quadrupole terms. For this case, there are no orbital flips, and very little change when $Q^2$ terms are turned on.    For Cases F and G, octupole terms now generate orbital flips, while $Q^2$ terms modify them somewhat, but do not suppress them.

\section{Discussion}
\label{sec:discussion}

We have derived the leading second-order quadrupole-quadrupole ($Q^2$) contributions to the secular evolution of hierarchical triple systems.  For systems where the mass of the third body is small compared to that of the inner binary, the effects, as expected, are unimportant.  But for systems where the third mass is larger than that of the inner binary, $Q^2$ effects can suppress orbital flips and extreme excursions of the inner eccentricity, that occur when the dynamics includes only first-order effects.  This suppression seems to occur when there is a fairly clear hierarchy between the dominant quadrupole timescale, the $Q^2$ timescale and the octupole timescale.  Table \ref{table:timescales} lists these timescales for the cases A through G, and the case LL studied in this paper.   For example, cases A, B and C show such a hierarchy of timescales, and all have flips suppressed cleanly by the $Q^2$ terms, while for case D, the  $Q^2$ and octupole timescales are comparable, and the result is a ragged ``Game of Thrones'' pattern of flips with the $Q^2$ terms turned on.   For case E, the octupole timescale is so large that there are no flips; for F and G, the octupole timescales are shorter, but the hierarchy of timescales and the weakening of $Q^2$ effects are such that a regular pattern of flips is preserved.  On the other hand, the case LL does not show the nice hierarchy of timecales (the $Q^2$ timescale is only twice the octupole timescale), yet the $Q^2$ terms cleanly suppress orbital flips.  For a more in-depth exploration of the presence or absence of orbital flips when these $Q^2$ effects are included, sec Sec.\ 4 of LKD \cite{2016MNRAS.458.3060L} .

The $Q^2$ effects we have disussed in this paper may have consequences for gravitational wave astronomy.  In the very high outer mass regime, corresponding to cases A, B and E, it appears that orbital flips and extreme eccentricities do not occur, either because $\epsilon$ is too small to produce significant octupole effects (case E), or because $\epsilon$ is so large that $Q^2$ effects suppress the flips (cases A and B).  This regime corresponds to binaries of $\sim 100 \, M_\odot$ total orbiting massive black holes  of $10^5 \, M_\odot$ and more.  Note that these conclusions are valid for arbitrary mass ratios in the inner binary.  In our numerical examples, we chose $m_1/m_2 = 0.01$ in order to enhance the flip-inducing octupole terms.  Increasing $m_1/m_2$ has no effect on our conclusions, and finally, when $m_1 =m_2$, octupole terms turn off completely, suppressing  flips independently of $Q^2$ terms.   On the other hand, while orbital flips and extreme eccentricities in this regime are suppressed, eccentricities as high as $0.99$ are routinely reached, leading to complex gravitational waveforms.

\begin{center}
\begin{table*}[t]
\caption{Relevant timescales for selected case studies} 
\begin{tabular}{l@{\hskip 0.5cm}c@{\hskip 0.5cm}c@{\hskip 0.5cm}c@{\hskip 0.5cm}c@{\hskip 0.5cm}c@{\hskip 0.5cm}c@{\hskip 0.5cm}c}
\hline
Case&$\alpha$&$\epsilon$&$T_{\rm Quad}$&$T_{Q^2}$&$T_{\rm Oct}$&$T_{\rm GR}$ \\
\hline 
A&$10^4$&$0.01$&$51$&$262$&$5460$&$5.3 \times 10^4$\\
B&$10^3$&$0.02$&$51$&$262$&$2530$&$5.3 \times 10^4$\\
C&$10^2$&$0.05$&$51$&$263$&$1180$&$5.3 \times 10^4$\\
D&$10$&$0.10$&$51$&$275$&$550$&$5.3 \times 10^4$\\
E&$10^4$&$0.003$&$1680$&$4.9 \times 10^4$&$5.7 \times 10^5$&$5.3 \times 10^6$\\
F&$10^2$&$0.015$&$1680$&$4.9 \times 10^4$&$1.2 \times 10^5$&$5.3 \times 10^6$\\
G&$10$&$0.031$&$1680$&$5.2 \times 10^4$&$5.7 \times 10^4$&$5.3 \times 10^6$\\
LL&$1$&$0.10$&$512$&$1.2 \times 10^4$&$5.5 \times 10^3$&$5.4 \times 10^6$\\
\hline
\end{tabular}
\label{table:timescales}
\end{table*}
\end{center}

\acknowledgments

This work was supported in part by the National Science Foundation,
Grant No.\  PHY 19-00188.  We are grateful to Adrian Hamers and Hanlun Lei for useful comments on a draft of this paper, We particularly thank Liantong Luo and Boaz Katz for their generous help demonstrating the equivalence of our results.    

\appendix

\section{Brief review of the two-timescale analysis}
\label{app:twotimescale}

In this Appendix we give a brief review of the two-timescale analysis used this paper.  This is a streamlined version of the description given in \cite{2017PhRvD..95f4003W}; see also \cite{1978amms.book.....B,1990PhRvD..42.1123L,2004PhRvD..69j4021M,2008PhRvD..78f4028H}.
We wish to consider the general set of first-order differential equations
\begin{equation}
\frac{d X_\alpha (t)}{dt} = \varepsilon Q_\alpha (X_\beta(t), t) \,.
\label{eqapp:dXdf}
\end{equation}
We anticipate that the solutions for the $X_\alpha$ will have pieces that vary on a ``short'' orbital time scale, corresponding to periodic functions of $t$, but may also have pieces that vary on a long time scale, of order $1/\varepsilon$ times the short time scale.   In a two-time-scale analysis \cite{1978amms.book.....B,1990PhRvD..42.1123L,2004PhRvD..69j4021M,2008PhRvD..78f4028H}, one treats these two times formally as independent variables, and solves the ordinary differential equations as if they were partial differential equations for the two variables.
We define the long-time-scale variable $\theta \equiv \varepsilon t$,
write the derivative with respect to $t$ as
${d}/{dt} \equiv \varepsilon {\partial}/{\partial \theta} + {\partial}/{\partial t}$
and define
\begin{equation}
X_\alpha (\theta, t) \equiv \tilde{X}_\alpha (\theta) + \varepsilon Y_\alpha (\tilde{X}_\beta (\theta), t) \,,
\label{eqapp:ansatz}
\end{equation}
where $\tilde{X}_\alpha (\theta)$ is the average of $X_\alpha$ over $t$, and $ Y_\alpha$ is the average-free part, 
where the average and average-free parts are defined by 
\begin{equation}
\langle A \rangle \equiv \frac{1}{T} \int_0^{T} A(\theta,t) dt \,,  \quad  {\cal AF}(A) \equiv  A(\theta,t) - \langle A \rangle \,, 
\label{eqapp:averagedef}
\end{equation}
where the integrals are carried out holding $\theta$ fixed.  

Substituting our definition of $X_\alpha$ into  Eq.\ (\ref{eqapp:dXdf}), and taking the average and average-free parts, we obtain the two main equations of the procedure
\begin{subequations}
\begin{align}
\frac{d\tilde{X}_\alpha}{d\theta} &= \langle Q_\alpha (\tilde{X}_\beta + \varepsilon Y_\beta, t) \rangle \,,
\label{eqapp:aveq}\\
\frac{\partial Y_\alpha}{\partial t} &= {\cal AF} \left (Q_\alpha (\tilde{X}_\beta + \varepsilon Y_\beta, t) \right )  - \varepsilon \frac{\partial Y_\alpha}{\partial \tilde{X}_\gamma} \frac{d\tilde{X}_\gamma}{d\theta} \,.
\label{eqapp:avfreeeq}
\end{align}
\label{eqapp:maineq}
\end{subequations}
Note that, by virtue of our assumption that $\theta$ and $t$ are independent, $\partial Y_\alpha/\partial \tilde{X}_\gamma$ is automatically average free.
Equation (\ref{eqapp:avfreeeq}) can be integrated, choosing the constant of integration so that the answer is average-free; the result is
\begin{align}
Y_\alpha (t)  &= {\cal AF} \left (\int_0^t \biggl [  {\cal AF} \left (Q_\alpha (\tilde{X}_\beta + \varepsilon Y_\beta, t) \right ) 
\right .
\nonumber \\
& \left .
\qquad \qquad
- \varepsilon \frac{\partial Y_\alpha}{\partial \tilde{X}_\gamma} \frac{d\tilde{X}_\gamma}{d\theta} \biggr ] dt' \right ) \,.
\end{align}

We now iterate Eqs.\ (\ref{eqapp:maineq}) in powers of $\varepsilon$.  We first expand
\begin{subequations}
\begin{align}
Y_\alpha  &\equiv Y^{(0)}_\alpha + \varepsilon Y^{(1)}_\alpha + O(\varepsilon^2 ) \,,
\label{eq:Yalphaexpand}
\\
Q_\alpha (\tilde{X}_\beta + \varepsilon Y_\beta, t) &\equiv Q_\alpha^{(0)} + 
\varepsilon Q_{\alpha,\beta}^{(0)} Y_\beta^{(0)} + O(\varepsilon^2)\,,
\label{eq:Qalphaexpand}
\end{align}
\end{subequations}
where
\begin{subequations}
\begin{align}
Y_\alpha^{(0)} &= {\cal AF} \left (\int_0^t {\cal AF} \left (Q_\alpha^{(0)} \right ) dt'  \right )\,,
\label{eq:Yalpha0}
\\
Q_\alpha^{(0)} &\equiv Q_\alpha (\tilde{X}_\beta, t) \,,
 \\
Q_{\alpha,\beta }^{(0)} & \equiv \frac{\partial Q_\alpha^{(0)} }{\partial \tilde{X}_\beta } \,.
\end{align}
\end{subequations}
To obtain $d\tilde{X}_\alpha/d\theta$ to order $\varepsilon^2$,  we substitute Eq.\ (\ref{eq:Yalpha0}) into Eq.\ (\ref{eq:Qalphaexpand}), convert back to the unscaled $t = \theta/\varepsilon$, and obtain Eq.\ (\ref{eq2:dXdtfinal}):
\begin{align}
\frac{d\tilde{X}_\alpha}{dt} &= \varepsilon \left \langle Q_\alpha^{(0)} \right \rangle + \varepsilon^2 \left  \langle {\cal AF} \left (Q_{\alpha,\beta}^{(0)} \right ) \int_0^t {\cal AF} \left (Q_{\beta}^{(0)} \right )dt' \right  \rangle
\nonumber \\
& \qquad \qquad
+ O(\varepsilon^3)\,,
\label{eqapp:dXdtfinal}
\end{align}
where we have employed the useful identity
\begin{align}
&\left \langle B \times {\cal AF} \left (\int_0^t {\cal AF} \left ( A \right ) dt' \right ) \right \rangle 
\nonumber \\
& \qquad \qquad
=  \left \langle  {\cal AF} \left (B \right ) \int_0^t {\cal AF} \left ( A \right ) dt' \right \rangle \,.
\end{align}
The first term in Eq.\ (\ref{eqapp:dXdtfinal}) is the standard first-order result in which ``constant'' values of the orbit elements are inserted into $Q_\alpha$ and the result is averaged over one period.  The second-order term results from the effect of periodic terms in the orbit elements on the behavior of the $Q_\alpha$.   Using the identity
\begin{equation}
\left \langle A \int_0^t B dt' \right \rangle = - \left \langle B \int_0^t A dt' \right \rangle + T\langle A \rangle\langle B \rangle \,,
\end{equation}
we can also express Eq.\ (\ref{eqapp:dXdtfinal}) in the equivalent form
\begin{align}
\frac{d\tilde{X}_\alpha}{dt} &= \varepsilon \left \langle Q_\alpha^{(0)} \right \rangle - \varepsilon^2 \left  \langle {\cal AF} \left (Q_{\beta}^{(0)} \right ) \int_0^t {\cal AF} \left (Q_{\alpha,\beta}^{(0)} \right )dt' \right  \rangle
\nonumber \\
& \qquad \qquad
+ O(\varepsilon^3)\,.
\label{eqapp:dXdtfinal2}
\end{align}

\section{The secular approximation in second-order perturbation theory}
\label{app:secular}

Because the application of the secular approximation to the quadrupole cross terms led to a result with the unexpected factor $P_{\rm out}/P_{\rm in}$ we will devote this Appendix to a detailed (if somewhat pedantic) study of this approximation.  The problem is to calculate a time average of combinations of periodic functions of time, one group of functions $A, \, B, \, \dots$ periodic with a period $P_1$, the other group of functions $M, \, N,\, \dots$  periodic with a period $P_2$, with $P_1 /P_2 \equiv \zeta \ll 1$.  The average is defined as 
\begin{equation}
\langle Q \rangle \equiv \frac{1}{T} \int_0^T Q(t) dt \,,
\label{eq:Qav}
\end{equation}
where $T$ is a suitably long time, say $nP_2$, where $n$ is an integer, yet still short compared with the timescale of the perturbations being analyzed.  

We will split this time into $m$ intervals of period $P_1$.  If $P_1$ and $P_2$ are not commensurate, there will be a fraction of a period $P_1$ left over.  However, we can choose $n$ and $m$ sufficiently large (subject to the timescale limitation mentioned above) so that the fractional mismatch of order $P_1/nP_2$ can be made smaller than some chosen tolerance.  Accordingly, to keep the calculation simple, we will assume that the periods are commensurate, so that $P_2/P_1 = m/n$.

We then break the integral in Eq.\ (\ref{eq:Qav}) into $m$ subintegrals of period $P_1$, to obtain
\begin{equation}
\langle Q(t) \rangle = \frac{1}{T} \sum_{q=0}^{m-1} \int_{qP_1}^{(q+1)P_1} Q(t) dt \,.
\end{equation}
Thus, for example, if $Q = A$, a function with periodicity $P_1$, the average becomes
\begin{align}
\langle A(t) \rangle &= \frac{1}{m} \sum_{q=0}^{m-1} \frac{1}{P_1}\int_{qP_1}^{(q+1)P_1} A(t) dt
\nonumber \\
&= \frac{1}{m} \sum_{q=0}^{m-1} \langle A \rangle_q
\nonumber \\
&= \frac{1}{P_1} \int_{0}^{P_1} A(t) dt \,,
\end{align} 
where we have used the fact that $\langle A \rangle_q$ is independent of $q$.  We note the useful fact that
\begin{align}
 \frac{1}{P_1} \int_{qP_1}^{(q+1)P_1} (t-qP_1)^n A(t)dt &= \frac{1}{P_1} \int_{0}^{P_1} t^n A(t)dt 
 \nonumber \\
 &= \langle t^n A(t) \rangle \,.
\end{align}

For the average of a long-period function $M(t)$, we assume that $M$ varies so slowly that we can Taylor expand $M(t)$ within each subinterval $q$.   This assumption ignores the phenomenon of {\em resonances}: if the outer orbit is eccentric, then there will be higher harmonics of the fundamental period, with periodicity $P_2/\ell$, where $\ell$ is an integer, and with amplitude decreasing as $E^\ell$.  If $P_2/\ell$ becomes comparable to the inner orbital period and the associated harmonic has sufficiently large amplitude, resonantly enhanced orbital perturbations can occur, often with striking consequences.  This is, of course, an entirely separate issue from the one we are exploring.  The standard secular approximation ignores resonances, and we will do so here.  Thus for the average of a function $M(t)$, we will write
\begin{align}
\langle M(t) \rangle &= \frac{1}{mP_1} \sum_{q=0}^{m-1} \int_{qP_1}^{(q+1)P_1} \biggl [
M_q 
  + (t-qP_1) \dot{M}_q 
  \nonumber \\
  & \qquad \qquad + O(t^2 \ddot{M}) \biggr ]dt \,,
\nonumber \\
&= \frac{1}{m} \sum_{q=0}^{m-1} M_q + \frac{P_1}{2m} \sum_{q=0}^{m-1} \dot{M}_q + O(\zeta^2 M) \,,
\label{eq:avM}
\end{align} 
where $M_q \equiv M(qP_1)$ and we recall that $\dot{M} \sim M/P_2$.
Now, since $M(t)$ is periodic with period $P_2$, the average of $\dot{M}$ vanishes, i.e.
\begin{align}
\langle \dot{M}(t) \rangle &= \frac{1}{mP_1} \left (M(nP_2) - M(0) \right ) 
 \nonumber \\
 &=0 
  \nonumber \\
 &=  \frac{1}{m} \sum_{q=0}^{m-1} \dot{M}_q + O(\zeta \dot{M}) \,.
\end{align}
Thus the second term in Eq.\ (\ref{eq:avM}) is of $O(\zeta^2 M)$ and we obtain
\begin{equation}
\langle M(t) \rangle = \frac{1}{m} \sum_{q=0}^{m-1} M_q + O(\zeta^2 M)  \,.
\end{equation}
Then the average of a product of functions $A(t)M(t)$ is given by
\begin{align}
\langle A(t) M(t) \rangle &= \frac{1}{mP_1} \sum_{q=0}^{m-1} \int_{qP_1}^{(q+1)P_1} A(t) 
\nonumber \\
& \qquad
\times  \biggl [ M_q 
  + (t-qP_1) \dot{M}_q + \dots \biggr ]dt \,,
\nonumber \\
&= \frac{1}{m} \langle A \rangle \sum_{q=0}^{m-1} M_q + \frac{1}{m} \langle tA(t) \rangle \sum_{q=0}^{m-1} \dot{M}_q + \dots \,,
\nonumber \\
&= \langle A \rangle \langle M \rangle + O(\zeta^2 AM) \,.
\end{align} 
This is the standard result in the secular approximation: the average of the products is equal to the product of the averages, up to corrections of order $\zeta^2$.

Using the same procedure, we can show that
\begin{subequations}
\begin{align}
\langle M(t) N(t) \rangle &= \frac{1}{m} \sum_{q=0}^{m-1} M_q N_q +  O(\zeta^2 MN) \,,
\label{eq:avMN}
 \\
\langle t N(t) \rangle &= \frac{P_1}{m} \sum_{q=0}^{m-1} \left ( q+\frac{1}{2} \right ) N_q
+ \frac{P_1^2}{2m} \sum_{q=0}^{m-1} q \dot{N}_q 
\nonumber \\
& \qquad
+ O(\zeta^2 P_2 N) \,,
\label{eq:avtM}
 \\
\langle t \dot{N}(t) \rangle &= \frac{P_1}{m} \sum_{q=0}^{m-1}  q  \dot{N}_q
 + O(\zeta N) \,.
 \label{eq:avtMdot}
\end{align}
\end{subequations}

\begin{widetext}
At second order in perturbation theory, we need to evaluate averages of integrals.  We begin with two simple examples.  Again breaking the integrals into subintegrals of size $P_1$, we obtain
\begin{subequations}
\begin{align}
\left \langle \int_0^t N(t') dt' \right \rangle &= \frac{1}{mP_1} \sum_{q=0}^{m-1}  \int_{qP_1}^{(q+1)P_1}  \biggl [ \, \sum_{r=0}^{q-1} \int_{rP_1}^{(r+1)P_1} N(t') dt' + \int_{qP_1}^{t} N(t') dt' \biggr ] dt 
\nonumber \\
& = - \frac{P_1}{m} \sum_{q=0}^{m-1} \left ( q+\frac{1}{2} \right ) N_q + P_1 \sum_{q=0}^{m-1} N_q - \frac{P_1^2}{2m} \sum_{q=0}^{m-1} q \dot{N}_q + O(\zeta^2 P_2 N) \,,
\label{eq:avintM}
\\
\left \langle M(t) \int_0^t N(t') dt' \right \rangle &=\frac{1}{mP_1} \sum_{q=0}^{m-1}  \int_{qP_1}^{(q+1)P_1}  M(t)  \biggl [ \, \sum_{r=0}^{q-1} \int_{rP_1}^{(r+1)P_1} N(t') dt' + \int_{qP_1}^{t} N(t') dt' \biggr ] dt 
\nonumber \\
& = \frac{P_1}{m} \sum_{q=0}^{m-1} \biggl [ M_q \sum_{r=0}^{q-1} N_r + \frac{1}{2} M_q N_q 
+ \frac{P_1}{2}  \sum_{r=0}^{q-1} \left (M_q \dot{N}_r + \dot{M}_q N_r \right ) \biggr ] +O(\zeta^2 P_2 MN) \,.
\label{eq:avMintN}
\end{align}
\end{subequations}
Note that Eqs.\ (\ref{eq:avtM}) and (\ref{eq:avintM}) satisfy the general result that
$\langle tQ \rangle + \langle \int Q \rangle = T \langle Q \rangle$.
We now want to evaluate the average $\langle AM \int BN \rangle$:
\begin{align}
\left \langle A(t) M(t) \int_0^t B(t') N(t') dt' \right \rangle &= \frac{1}{mP_1} \sum_{q=0}^{m-1}  \int_{qP_1}^{(q+1)P_1} A(t) \biggl [ M_q + (t-qP_1)\dot{M}_q + \dots \biggr ] dt
\nonumber \\
& \qquad 
\times \biggl \{  \, \sum_{r=0}^{q-1} \int_{rP_1}^{(r+1)P_1} B(t') \biggl [ N_r + (t' - rP_1)\dot{N}_r + \dots \biggr ] dt'
\nonumber \\
& \qquad \qquad
+\int_{qP_1}^{t} B(t') \biggl [ N_q + (t' - qP_1)\dot{N}_q + \dots \biggr ] dt'
\biggr \} \,,
\nonumber \\
&=\frac{1}{m} \sum_{q=0}^{m-1} \biggl [ P_1 \langle A \rangle \langle B \rangle  M_q \sum_{r=0}^{q-1} N_r + \left \langle A \int_0^t B dt' \right \rangle   M_q N_q
+P_1 \langle A \rangle \langle tB \rangle  M_q \sum_{r=0}^{q-1} \dot{N}_r
\nonumber \\
& \qquad
+ P_1 \langle tA \rangle \langle B \rangle  \dot{M}_q \sum_{r=0}^{q-1} N_r \biggr ]
+O(\zeta^2 P_2 AMBN)
  \,.
 \label{eq:avAMintBN}
\end{align}
The third and fourth terms in Eq.\ (\ref{eq:avAMintBN}) can be simplified by using Eq.\ (\ref{eq:avMintN}) to evaluate
$\langle M \int \dot{N} dt' \rangle$ and $\langle \dot{M} \int {N} dt' \rangle$ to the leading order in $\zeta$, to obtain
\begin{align}
\frac{P_1}{m} \sum_{q=0}^{m-1} M_q \sum_{r=0}^{q-1} \dot{N}_r &= \langle MN \rangle - N(0) \langle M \rangle + O(\zeta P_2 MN)\,,
\nonumber \\
\frac{P_1}{m} \sum_{q=0}^{m-1} \dot{M}_q \sum_{r=0}^{q-1} {N}_r &=- \langle MN \rangle + M(0) \langle N \rangle + O(\zeta P_2 MN) \,.
\label{eq:sumsimplify}
\end{align}
Thus Eq.\ (\ref{eq:avAMintBN}) becomes
\begin{align}
\left \langle A(t) M(t) \int_0^t B(t') N(t') dt' \right \rangle &= \langle A \rangle \langle B \rangle \left \langle M \int_0^t N dt' \right \rangle 
+\left ( \left \langle A \int_0^t B dt' \right \rangle - \frac{1}{2} P_1 \langle A \rangle \langle B \rangle \right ) \langle MN \rangle
\nonumber \\
& \qquad
+  \langle A  \rangle \left \langle \left ( t -  \frac{1}{2} P_1 \right ) B \right \rangle \left (\langle MN \rangle - N(0) \langle M \rangle \right )
\nonumber \\
& \qquad
- \langle B \rangle \left \langle \left ( t -  \frac{1}{2} P_1 \right ) A \right \rangle  \left (\langle MN \rangle - M(0) \langle N \rangle \right ) +O(\zeta^2 AMBN)\,.
\end{align}
From this it is then straightforward to show that, for average-free quantities,
\begin{align}
\left \langle {\cal AF}(A(t) M(t)) \int_0^t {\cal AF} (B(t') N(t')) dt' \right \rangle &= \langle A \rangle \langle B \rangle \left \langle{\cal AF} (M) \int_0^t {\cal AF} (N) dt' \right \rangle 
+ \left \langle {\cal AF} (A) \int_0^t {\cal AF}  (B) dt' \right \rangle  \langle MN \rangle
\nonumber \\
& \qquad
+ O(\zeta^2 P_2 AMBN) \,.
\label{eq:avfinal}
\end{align}
Because of the time integrals, the first term in Eq.\ (\ref{eq:avfinal}) is of order $P_2$ times $AMBN$, the second term is of order $P_1$ or $\zeta P_2$ times $AMBN$; we ignore subdominant terms of order $\zeta^2 P_2$ or $\zeta P_1$ times $AMBN$.   This unexpected enhancement by the factor $P_2$ occurs only when both functions involved in the average of a second-order term involve two orbital timescales.  If, for example, the second-order term is a cross term between a post-Newtonian perturbation of the inner orbit and a multipolar perturbation caused by the outer body, then either $M = 1$ or $N = 1$, and the first term in Eq.\ (\ref{eq:avfinal}) vanishes.  In these cases, we obtain the expected average
\begin{align}
\left \langle {\cal AF}(A(t)) \int_0^t {\cal AF} (B(t') N(t')) dt' \right \rangle 
 = 
 \left \langle {\cal AF} (A) \int_0^t {\cal AF}  (B) dt' \right \rangle  \langle N \rangle
+ O(\zeta P_1 AMBN) \,,
\end{align}
with an analogous result for $N = 1$.  We will see these averages at work in future papers \cite{willcrossterms}.

\section{Comparison with the results of LKD}
\label{app:Luo}

Luo, Katz and Dong \cite{2016MNRAS.458.3060L} developed an approach called ``Corrected Double Averaging'' (CDA) to go beyond the standard application of the secular approximation at first order in perturbation theory.  This approach explicitly takes into account the periodic perturbations of the orbit  before averaging over the two orbital timescales.  They work in terms of equations of motion expanded to quadrupole order for the normalized angular momentum vector $\bm{j} = \bm{h} /\sqrt{Gma}$ and the Runge-Lenz vector $\bm{A} = \bm{v} \times \bm{h} /Gm - \bm{n}$ for the inner orbit.  They use equations for $d\bm{j}/dt$ and $d\bm{e}/dt$ that have already been averaged over the inner orbit, which is equivalent to singling out the effects of ``term 1'' in Eq.\ (\ref{eq:secularnew}).   They then find the solutions periodic in $F$ (plus terms linear in $F$), reinsert them into the equations of evolution and average over $F$.  The results are displayed in Eqs.\ (C1) and (C2) of  \cite{2016MNRAS.458.3060L}.

Those equations can be seen to be completely equivalent to Eqs.\ (\ref{eq:QQterms}) by making the following change of variables from our $\tilde{X}_\alpha$ to the corresponding variables $X_\alpha^{\rm LKD}$ inferred from the components of $\bm{j}$ and $\bm{e}$ in \cite{2016MNRAS.458.3060L}:
\begin{equation}
X_\alpha^{\rm LKD} = \tilde{X}_\alpha + {\cal K} \delta X_\alpha \,,
\end{equation}
where (dropping the tildes)
\begin{align}
\delta e &=  \frac{5}{16}  \frac{\alpha \epsilon^{3/2}}{(1+\alpha)^{1/2}} e(1-e^2)^{1/2}
\biggl ((1+\cos z)^2 \cos(2\omega-2\omega_3)
-(1-\cos z)^2 \cos(2\omega+2\omega_3) \biggr ) \,,
\nonumber \\
\delta z &= \frac{1}{16}  \frac{\alpha \epsilon^{3/2}}{(1+\alpha)^{1/2}} \frac{ \sin z}{(1-e^2)^{1/2}}
\biggl (5e^2(1+\cos z) \cos(2\omega-2\omega_3)
 +5e^2(1-\cos z) \cos(2\omega+2\omega_3) 
+2 (2+3e^2) \cos 2\omega_3  \biggr ) \,,
\nonumber \\
\delta \Omega &= \frac{1}{16}  \frac{\alpha \epsilon^{3/2}}{(1+\alpha)^{1/2}} \frac{1}{(1-e^2)^{1/2}}
\biggl (5e^2(1+\cos z) \sin(2\omega-2\omega_3)
 +5e^2(1-\cos z) \sin(2\omega+2\omega_3) 
+2 (2+3e^2) \cos z \sin 2\omega_3  \biggr ) \,,
\nonumber \\
\delta \omega &= -\frac{1}{16}  \frac{\alpha \epsilon^{3/2}}{(1+\alpha)^{1/2}} (1-e^2)^{1/2} 
\biggl (5(1+\cos z)^2 \sin(2\omega-2\omega_3)
 -5(1-\cos z)^2 \sin(2\omega+2\omega_3) 
-6 \sin^2 z \sin 2\omega_3  \biggr )
\nonumber \\
& \qquad \quad  - \cos z \, \delta \Omega \,,
\label{eq:transform}
\end{align}
where
\begin{equation}
{\cal K} = \frac{1}{2} \frac{E^2}{(1-E^2)^{3/2}}\frac{ (1+ 2\sqrt{1-E^2} )}{ (1+ \sqrt{1-E^2} )^2}\,.
\end{equation}
Subsequently, LKD spotted a subtlety in how expressions linear in $F$ were to be averaged over the outer orbit in the CDA method, leading to a transformation from the original averaged $\bm{j}$ and ${\bm e}$ to a new pair of vectors.  Those transformations are precisely the same as Eq.\ (\ref{eq:transform}).  As a result, the two methods are in complete agreement.

\end{widetext}


\end{document}